\documentclass[preprint,3p,times]{elsarticle}
\usepackage{graphicx}
\usepackage{amssymb}
\usepackage{natbib}
\usepackage{textcomp}
\usepackage{amsmath, amsthm}
\usepackage{subfig}
\usepackage{color}
\journal{Journal of Computational Physics}

\begin{document}

\begin{frontmatter}


%%\tnotetext[label2]{}
\fntext[email]{Tel.: +44 (0)1895 267371 \\
                  E-mail: Jan.wissink@brunel.ac.uk}

\title{Low-diffusivity scalar transport using a WENO scheme and dual meshing}

\address[brunel]{Brunel University London, Kingston Lane, Uxbridge, UB8 3PH, UK}
\address[KIT]{KIT, Karlsruhe Institute of Technology, Germany}
\address[dresden]{Technical University Dresden, Germany}

\author[brunel]{B.~Kubrak}
\author[KIT]{H.~Herlina}
\author[dresden]{F.~Greve}
\author[brunel]{J.G.~Wissink\fnref{email}}

\address{}

\begin{abstract}

Interfacial mass transfer of low-diffusive substances in an unsteady flow environment is marked by a very thin boundary layer at the interface and other regions 
with steep concentration gradients. A numerical scheme capable of resolving accurately most details of this process is 
presented. In this scheme, the fifth-order accurate WENO method developed by \citet{Liu_osher_chan_94} was implemented 
on a non-uniform staggered mesh to discretize the scalar convection while for the scalar diffusion a fourth-order
accurate central discretization was employed. The discretization of the scalar convection-diffusion equation 
was combined with a fourth-order Navier-Stokes solver which solves the incompressible flow. 
A dual meshing strategy was employed, in which the scalar was solved on
a finer mesh than
the incompressible flow.
The order of accuracy of the solver for one-dimensional scalar transport was tested
on both stretched and uniform grids. Compared
to the fifth-order WENO implementation of \citet{Jiang_Shu_96}, the
\citet{Liu_osher_chan_94} method was found to be superior on very coarse
meshes.
The solver was further tested by performing a number of two-dimensional
simulations.  At first a grid refinement test was performed at zero
viscosity with shear acting on an initially axisymmetric scalar
distribution. A second refinement test was conducted for an unstably
stratified flow with low
diffusivity scalar transport. The unstable stratification led
to buoyant convection which was modelled using a
Boussinesq approximation with a linear relationship between flow
temperature and density. The results show that for
the method presented a
relatively coarse mesh is sufficient to accurately describe the fluid
flow, while the use of a refined dual mesh for the
low-diffusive scalars is found to be beneficial in order to obtain a
highly accurate resolution with negligible numerical diffusion.

\end{abstract}

\begin{keyword}

Air-Water Interface, DNS, Gas Transfer, WENO Scheme, Scalar Transport, high Schmidt number. 
\end{keyword}

\end{frontmatter}

% \linenumbers

\section{Introduction}
To accurately resolve low-diffusivity scalar transport problems special numerical schemes are necessary for 
the discretization of the convective term in order to avoid under- and/or overshoots of the scalar quantity. The first order upwind 
method, for example, could be used to effectively avoid such under- and/or overshoots but at the cost of introducing an excessive 
amount of numerical diffusion \cite{Collatz_Book, Sharif_1987}.
Up to now, a number of DNS studies of gas transfer across the air-water interfaces have been carried out for shear driven and 
stirred vessels.  Hasegawa and Kasagi \cite{Hasegawa_2009}
studied wind-shear driven mass transfer across the turbulent interface at a Schmidt number of $Sc=100$.
They used a pseudo-spectral Fourier method for the spatial discretization
in the horizontal directions, whereas the finite volume method is employed in the normal direction in which turbulent and molecular 
mass fluxes are evaluated at a cell surface with second-order accuracy.
Handler et al. \cite{Handler_1999} used a pseudo spectral approach with Fourier expansions to carry out direct numerical simulations 
for the transport of a passive scalar at a shear-free boundary in fully developed channel flow. Similarly, Banerjee et al.
also used a pseudo-spectral method to extensively study the mechanisms of turbulence and scalar exchange at 
the air-water interface in several publications (see \cite{Banerjee_2004, Banerjee_2007} and references therein).
Schwertfirm and Manhart \cite{Schwertfirm_2007} also studied passive scalar transport in a turbulent channel flow for Schmidt numbers up
to $Sc=49$. 
They used a similar approach as presented in the present work by solving the scalar on a finer grid than the velocity which was mapped by 
a conservative interpolation to the fine-grid. An explicit iterative finite-volume scheme of sixth-order accuracy was employed 
to calculate all convective and diffusive fluxes, while for the time-integration a third order Runge-Kutta method was used \cite{Schwertfirm_2008}.

The pseudo-spectral methods used above have excellent error properties when the solution is relatively smooth. However, a main 
disadvantage of using spectral methods lies in the formation of
non-physical oscillations near steep gradients that may regularly occur in the solution of a convection-diffusion problem 
when the diffusivity is extremely small. These Gibbs oscillations (under- and/or overshoots) \cite{Tadmor_1990} near steep gradients 
are not uncommon and can also be found when higher-order central finite-difference methods are used on a relatively coarse mesh 
to discretize the convection of a scalar with low diffusivity.
This can be overcome by using weighted essentially non-oscillatory schemes (WENO) which have excellent shock capturing capabilities. 
Their non-oscillatory behaviour is advantageous in dealing with very steep gradients. In this paper, we present a numerical scheme 
developed specifically to resolve the details of the interfacial mass transfer of low-diffusive substances by adapting the weighted 
essentially non-oscillatory (WENO) by Liu et al. \cite{Liu_osher_chan_94}.
They are based on essentially non-oscillatory (ENO) schemes which were first published
in the meanwhile classic paper of Harten et al. \cite{Harten_87}. Liu et al. \cite{Liu_osher_chan_94} introduced 
the idea of taking a convex combination of interpolation polynomials to construct a stencil using non-linear weights with a high 
order-of-accuracy in smooth regions while weighing out the non-smooth stencils in regions containing steep gradients or discontinuities.
They studied WENO($2r-1$) schemes for different stencil sizes, i.e. $r=2$ (WENO3) and $r=3$ (WENO5). 

In the meantime a large variety of WENO schemes has been developed. Many improvements were made 
by modifying the smoothness determination. For instance, Jiang and Shu \cite{Jiang_Shu_96} introduced a new smoothness
indicator that is used to evaluate the non-linear weights. The size of the stencil has also been further increased by 
Balsara and Shu \cite{Balsara_Shu_2000} extending it up to $r=6$ (WENO11). 
Henrick et al. \cite{Henrick_2005} could show that the weights generated by the classical choice of smoothness indicators in \cite{Jiang_Shu_96}
failed to recover the maximum
order of the scheme at critical points of the solution where the first derivatives are zero.
They developed the so called 
WENOM schemes where a mapping procedure is introduced to keep the weights of the stencils as close as possible to the optimal weights.
The resulting (mapped) WENOM scheme of \citet{Henrick_2005} presented more accurate results close to discontinuities. 
Even more recently, \citet{Borges_2008} achieved the same results as mapped WENO schemes without mapping but 
by improving the accuracy of the classical WENO5 scheme by devising a new smoothness indicator and non-linear weights using 
the whole 5-points stencil and not the classical smoothness indicator of \citet{Jiang_Shu_96} which
uses a composition of three 3-points stencils. 
The schemes of \citet{Borges_2008} are known as WENO-Z schemes. Of all the schemes discussed above the classical WENO5 scheme is used most widely 
\cite{Henrick_2005, Martin_Taylor_2006, Johnson_2006}. 
In our simulations we do not expect any discontinuities in the scalar field so that the classical WENO5 scheme of Liu et al. \cite{Liu_osher_chan_94}
is a good choice to accurately resolve low-diffusive scalar transport which may lead to steep concentration gradients. 

An example of application is given for the 2D case of buoyant-convectively 
driven mass transfer with a Prandtl number of $Pr =6$ and a Schmidt number 
of $Sc = 500$. 
One typical process in nature of such a case is the absorption 
of oxygen into lakes during night time. This process is controlled by the low 
diffusivity of the dissolved gas in the water 
and the convective instability triggered by the density difference 
between the cold water at the top surface and the warm water in the bulk. 
The convective-instability enhances the gas transfer into the 
water body significantly compared with the static condition with only diffusive 
gas transfer. This low diffusive process results in a very thin concentration 
boundary layer that is found at the water surface. Experimental measurements 
near the surface (such as the mass flux) are very difficult. 
There is a need to fully resolve the near surface mechanisms in order to 
understand the physical mechanism.

Below, the capability of the newly developed code to accurately resolve 
low-diffusivity scalar transport problems will be illustrated. 
First the full set of 2D equations to be solved and the formulation of the 
numerical schemes used in Section~\ref{sec:numericalmethod} are presented. 
Section~\ref{sec:1DNumericalExperiments} covers 1D numerical experiments that 
were performed to determine the accuracy of the applied schemes on uniform 
and stretched meshes for both purely convective and purely diffusive scalar 
transport. 
The numerical schemes were further tested by performing a number of 
2D-simulations. The first 2D application case, presented in Section 4, 
deals with a zero-viscosity steady shear flow acting on an initially 
axisymmetric scalar distribution. 
In the last section the solver was tested for the 2D case of low-diffusivity 
scalar transport in buoyancy driven flow. 

\section{Formulation of Numerical Method}
\label{sec:numericalmethod}
The full set of the 2D governing equations to be solved in order to 
address/simulate the low-diffusivity scalar transport problem are first 
presented in this section followed by the formulation of the numerical schemes 
used. 
For the scalar transport the two-dimensional convection diffusion equation 
of the scalar $\varphi=\varphi(x,z,t)$ in conservative form reads
\begin{equation}
 \frac{\partial \varphi}{\partial t}+\frac{\partial u \varphi}{\partial x}+
 \frac{\partial w \varphi}{\partial z}=
D  \left(\frac{\partial^{2} \varphi}{\partial x^{2}}+\frac{\partial^{2} \varphi}{\partial z^{2}}\right),
\label{eq:conv_diff}
\end{equation}
where $x$ and $z$ are the horizontal and vertical directions, respectively, 
$u$ and $w$ are the velocities in the $x$ and $z$ directions, $D$ is the molecular diffusion coefficient of the dissolved substance and $t$ denotes 
time.

For the flow-field the incompressible Navier-Stokes equation is solved. The continuity equation for two-dimensional incompressible flow reads,
\begin{equation}
 \frac{\partial u}{\partial x}+\frac{\partial w}{\partial z}=0,
 \label{eqn:continuity}
\end{equation}
and the momentum equations are given by
\begin{eqnarray}
 \frac{\partial u}{\partial t}&=&-\frac{\partial p}{\partial x}
 + a \label{eqn:Naw_stoke_x} \\
 \frac{\partial w}{\partial t}&=&-\frac{\partial p}{\partial z}
 + c  \label{eqn:Naw_stoke_y}
\end{eqnarray}
where $p$ is pressure and $a$ and $c$ represent the sum of the convective and diffusive terms
\begin{eqnarray}
a&=&-\frac{\partial u^{2}}{\partial x}-\frac{\partial uw}{\partial z}
 + \frac{1}{Re}\left\{\frac{\partial^2 u}{\partial x^{2}}+\frac{\partial^2 u}{\partial z^{2}}\right\} \label{eqn:a_momentum} \\
c&=&-\frac{\partial w^{2}}{\partial z}-\frac{\partial uw}{\partial x}
 + \frac{1}{Re}\left\{\frac{\partial^2 w}{\partial x^{2}}+\frac{\partial^2 w}{\partial z^{2}}\right\} %+\beta (T^*). 
 \label{eqn:c_momentum}
\end{eqnarray}
where $Re$ is the Reynolds number.

\subsection{Discretization of the convection-diffusion equation of the scalar $\varphi$}
%%

%The schemes that are employed here are variants of the WENO5 finite difference scheme as 
%described by \citet{Liu_osher_chan_94} and \citet{Jiang_Shu_96}. 

In this section we outline the discretization of the transport equation  for the scalar $\varphi$ as given in equation (\ref{eq:conv_diff}).
%The two-dimensional convection diffusion equation
%of $\varphi=\varphi(x,z,t)$ reads 
%
%\begin{equation}
% \frac{\partial \varphi}{\partial t}+u\frac{\partial \varphi}{\partial x}+w\frac{\partial \varphi}{\partial z}=
%\Gamma  \left(\frac{\partial^{2} \varphi}{\partial x^{2}}+\frac{\partial^{2} \varphi}{\partial z^{2}}\right),
%\label{eq:conv_diff}
%\end{equation}
%where $x$ and $z$ are the horizontal and vertical directions, respectively and $t$ denotes time.
The diffusive term on the right of (\ref{eq:conv_diff}) is discretized using a fourth-order accurate central scheme, while the convective term is discretized
using variants of the fifth-order WENO schemes developed by \citet{Liu_osher_chan_94} and \citet{Jiang_Shu_96}. The WENO schemes use an approximation of the scalar fluxes at the cell interface by employing 
interpolation schemes. The reconstruction procedure produces a high order accurate approximation
of the solution from the calculated cell averages.
Below the
implemented scheme is detailed only in one dimension. Generalization to 
higher dimensions is straightforward.

When ignoring the diffusive term, the one dimensional variant 
of \eqref{eq:conv_diff} can be rewritten as
\begin{equation}
 \frac{\partial \varphi}{\partial t}=-\frac{\partial u \varphi}{\partial x} 
\end{equation}
where $u$ is the velocity in the $x$-direction.
As we employ a staggered mesh, for the volume centred around $x=x_{i}$, 
the convective fluxes $R_{i}^{+}$ and $R_{i}^{-}$
are defined by

\begin{eqnarray}
R_{i}^{+}&=&\frac{a_{0}}{a_{0}+a_{1}+a_{2}}P_{i-1}(x_{i+\frac{1}{2}})+\frac{a_{1}}{a_{0}+a_{1}+a_{2}}P_{i}(x_{i+\frac{1}{2}})
+\frac{a_{2}}{a_{0}+a_{1}+a_{2}}P_{i+1}(x_{i+\frac{1}{2}})
\label{eq:Rplus}
\end{eqnarray}
and
\\
\begin{eqnarray}
R_{i}^{-}&=&\frac{a_{0}}{a_{0}+a_{1}+a_{2}}P_{i-1}(x_{i-\frac{1}{2}})+\frac{a_{1}}{a_{0}+a_{1}+a_{2}}P_{i}(x_{i-\frac{1}{2}})
+\frac{a_{2}}{a_{0}+a_{1}+a_{2}}P_{i+1}(x_{i-\frac{1}{2}})
\label{eq:Rminus}
\end{eqnarray}
where $a_0, a_1, a_2$ are coefficients and $P$ the Lagrange interpolations polynomials defined in \eqref{eqn:P_lagrange}.
In the implementation of \citet{Liu_osher_chan_94}, for $R_{i}^{+}$ the weights for the convex combination of the quadratic Lagrange interpolation polynomials are given by,
\begin{equation}
 a_{0}=\frac{1}{12(\varepsilon + IS_{i})^3},\;\;
 a_{1}=\frac{1}{2(\varepsilon + IS_{i+1})^3},\;\;
 a_{2}=\frac{1}{4(\varepsilon + IS_{i+2})^3},
 \label{eq:aplus_LiuOsherChan}
\end{equation}
\noindent while for $R_{i}^{-}$ the weights are given by
\begin{equation}
a_{0}=\frac{1}{4(\varepsilon + IS_{i})^3},\;\;
 a_{1}=\frac{1}{2(\varepsilon + IS_{i+1})^3},\;\;
 a_{2}=\frac{1}{12(\varepsilon + IS_{i+2})^3},
 \label{eq:aminus_LiuOsherChan}
\end{equation}
where $\varepsilon=10^{-6}$ and the smoothness indicator $IS_{i}$ is defined by
\begin{eqnarray}
IS_{i}=\frac{1}{2}((\varphi_{i-1}-\varphi_{i-2})^{2}+(\varphi_{i}-\varphi_{i-1})^2)
+(\varphi_{i}-2\varphi_{i-1}+\varphi_{i-2})^{2}.
\label{eq:IS_LiuOsherChan}
\end{eqnarray}

The original calculation of the weights $a_{0}$, $a_{1}$ and $a_{2}$ as presented above is compared to an alternative developed by \citet{Jiang_Shu_96}, in which the weights for $R_{i}^{+}$ are given by
\begin{equation}
 a_{0}=\frac{1}{10(\varepsilon + IS_{0})^r},\;\;  
 a_{1}=\frac{6}{10(\varepsilon + IS_{1})^r},\;\;  
 a_{2}=\frac{3}{10(\varepsilon + IS_{2})^r}, 
 \label{eq:aplus_JiangShu}
\end{equation}
\noindent while for $R_{i}^{-}$ the weights are given by
\begin{equation}
 a_{0}=\frac{3}{10(\varepsilon + IS_{0})^r},\;\; 
 a_{1}=\frac{6}{10(\varepsilon + IS_{1})^r},\;\;
 a_{2}=\frac{1}{10(\varepsilon + IS_{2})^r},
 \label{eq:aminus_JiangShu}
\end{equation}
with the smoothness indicators $IS_{i}$ defined by
\begin{eqnarray}
 IS_{0}&=&\frac{13}{12}(\varphi_{i-2}-2\varphi_{i-1}+\varphi_{i})^{2}
+\frac{1}{4}(\varphi_{i-2}-4\varphi_{i-1}+3\varphi_{i})^{2} \nonumber \\
 IS_{1}&=&\frac{13}{12}(\varphi_{i-1}-2\varphi_{i}+\varphi_{i+1})^{2}
+\frac{1}{4}(\varphi_{i-1}-\varphi_{i+1})^{2} \nonumber \\
 IS_{2}&=&\frac{13}{12}(\varphi_{i}-2\varphi_{i+1}+\varphi_{i+2})^{2}
+\frac{1}{4}(3\varphi_{i}-4\varphi_{i+1}+\varphi_{i+2})^{2}.
\label{eq:IS_JiangShu}
\end{eqnarray}
Note that in the 1D tests presented below a power of $r=3$ 
in~(\ref{eq:aplus_JiangShu}) and~(\ref{eq:aminus_JiangShu}) was used.
The upstream central method can be obtained from both the \citet{Liu_osher_chan_94} 
and \citet{Jiang_Shu_96} implementations of the weights by setting all 
smoothness indicators $IS$ in (\ref{eq:aplus_LiuOsherChan}), 
(\ref{eq:aminus_LiuOsherChan}) and (\ref{eq:aplus_JiangShu}), 
(\ref{eq:aminus_JiangShu}) to zero.

The modified quadratic Lagrange interpolations $P_{i}(x)$ in 
equations \eqref{eq:Rplus} and \eqref{eq:Rminus} read
\begin{eqnarray}
 P_{i}(x)=&\frac{(x-x_{i})(x-x_{i+1})}{(x_{i-1}-x_{i})(x_{i-1}-x_{i+1})}\varphi_{i-1} 
 +\frac{(x-x_{i-1})(x-x_{i+1})}{(x_{i}-x_{i-1})(x_{i}-x_{i+1})}\varphi_{i}
+\frac{(x-x_{i-1})(x-x_{i})}{(x_{i+1}-x_{i-1})(x_{i+1}-x_{i})}\varphi_{i+1} \nonumber \\
&-\frac{(x_{i}-x_{i-1})\varphi_{i+1}-(x_{i+1}-x_{i-1})\varphi_{i}+(x_{i+1}-x_{i})\varphi_{i-1}}{12(x_{i+1}-x_{i})}
\label{eqn:P_lagrange}
\end{eqnarray}

\begin{figure}[h!]
 \centering
 \includegraphics[width=0.45\textwidth]{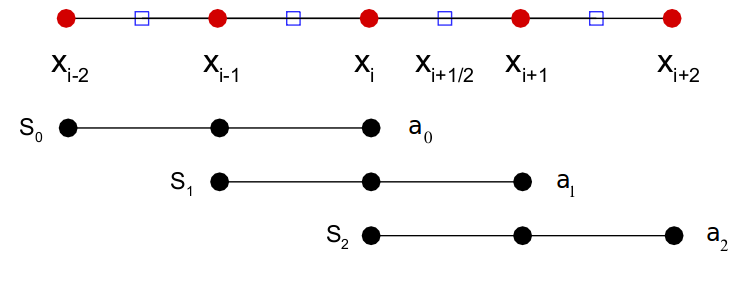}
 \caption{Schematic illustration of the weighted 5 point convex combination composed of three 3-points stencils $S_{0},S_{1},S_{2}$
  and their respective weights $a_{0},a_{1},a_{2}$ used in the classical WENO5 scheme. \cite{Shu_Osher_88}.}
 \label{fig:stencils}
\end{figure}

High-order polynomial interpolations to the midpoints $x_{i+\frac{1}{2}}$ are computed 
using known grid values of the scalar $\varphi$. The scheme uses a 5-points stencil which is divided into three 3-points 
stencils as shown in Fig.~\ref{fig:stencils}.

These are interpolations of the scalar to the faces of the volume combined with a smoothing term at the right.
Using the above, depending on the signs of $u_{i-\frac{1}{2}}$ and $u_{i+\frac{1}{2}}$, we have four possible ways to
calculate the discretization of the convective terms  $L_{i}(\varphi)~=~\left(-u\frac{\partial\varphi}{\partial x}\right)|_{x_{i}}$ in $x_i$:

\begin{eqnarray}
 u_{i+\frac{1}{2}}>0,u_{i-\frac{1}{2}}>0:L_{i}(\varphi)&=&-\frac{u_{i+\frac{1}{2}}R^{+}_{i}-u_{i-\frac{1}{2}}R^{+}_{i-1}}{x_{i+\frac{1}{2}}-x_{i-\frac{1}{2}}} \nonumber \\
 u_{i+\frac{1}{2}}>0,u_{i-\frac{1}{2}}<0:L_{i}(\varphi)&=&-\frac{u_{i+\frac{1}{2}}R^{+}_{i}-u_{i-\frac{1}{2}}R^{-}_{i}}{x_{i+\frac{1}{2}}-x_{i-\frac{1}{2}}} \nonumber \\
 u_{i+\frac{1}{2}}<0,u_{i-\frac{1}{2}}>0:L_{i}(\varphi)&=&-\frac{u_{i+\frac{1}{2}}R^{-}_{i+1}-u_{i-\frac{1}{2}}R^{+}_{i-1}}{x_{i+\frac{1}{2}}-x_{i-\frac{1}{2}}} \nonumber \\
 u_{i+\frac{1}{2}}<0,u_{i-\frac{1}{2}}<0:L_{i}(\varphi)&=&-\frac{u_{i+\frac{1}{2}}R^{-}_{i+1}-u_{i-\frac{1}{2}}R^{-}_{i}}{x_{i+\frac{1}{2}}-x_{i-\frac{1}{2}}} \nonumber \\
\nonumber
\end{eqnarray}
By examining the equations above it can be seen that at each interface 
$x_{i+\frac{1}{2}}$ of 2 neighbouring cells the scalar flux 
(either $u_{i+\frac{1}{2}}R^+_i$ or $u_{i+\frac{1}{2}}R^-_{i+1}$) 
is uniquely determined, which ensures that any scalar quantity that leaves the volume
centred around $x_i$ through this interface will enter the
volume centred around $x_{i+1}$.
Upwind information is incorporated by the way in which the scalar at each cell 
interface is interpolated. 
For instance, the interpolation stencil for $R^+_i$ (\ref{eq:Rplus}) - which 
is employed when $u_{i+\frac{1}{2}} > 0$ and consists of five points with 
$x=x_i$ in the middle - is used to calculate the scalar at the 
location $x=x_{i+\frac{1}{2}}$ which is located upstream (upwind) of $x=x_i$. 
A similar argument holds for the calculation of $R^-_i$ (\ref{eq:Rminus}). Hence,
both stencils are non-symmetric and use more information from the upwind direction than
from the downwind direction. Based on
this bias, the method discussed above can be classified as an upwind method. \\
\noindent
With the methods described here a fifth-order accuracy can be achieved.
Note that the weights given to the interpolating polynomials $a_0, a_1, a_2$  depend on the local smoothness of the solution. 
Interpolation polynomials defined in regions where
the solution is smooth are given higher weights than those in regions near discontinuities (shocks) or 
steep gradients (like the gas concentration near the interface in our application case presented in Section~\ref{sec:Mesh_sensitivity_2D_buoyancy}).

The diffusive term on the right hand side of \eqref{eq:conv_diff} is discretized using a fourth-order central finite difference method 
for the second derivative such as,
\begin{equation}
 \frac{\partial^{2} \varphi}{\partial x^{2}} \approx \frac{-\varphi_{i+2,k}+16\varphi_{i+1,k}-30\varphi_{i,k}+16\varphi_{i-1,k}-\varphi_{i-2,k}}{12 (\delta x_i)^2}
\label{eqn:diffusion_discretize_x}
\end{equation}
and
\begin{equation}
 \frac{\partial^{2} \varphi}{\partial z^{2}} \approx \frac{-\varphi_{i,k+2}+16\varphi_{i,k+1}-30\varphi_{i,k}+16\varphi_{i,k-1}-\varphi_{i,k-2}}{12(\delta z_k)^2}
\label{eqn:diffusion_discretize_z}
\end{equation}
where $\delta x_i = x_{i+\frac{1}{2}}-x_{i-\frac{1}{2}}$ and $\delta z_k = z_{k+\frac{1}{2}}-z_{k-\frac{1}{2}}$, respectively. 
On a stretched mesh the actual discretization coefficients are obtained from the above equations using Lagrange interpolations to a seven-point numerical stencil.
The time integration of the convection-diffusion equation 
is implemented using a third order Runge-Kutta method (RK3) developed by Shu and Osher \cite{Shu_Osher_88} that reads,

\begin{eqnarray}
\varphi_{i}^{(1)}&=&\varphi_{i}^{n}+\Delta t L_{i}(\varphi_{i}^{n}) \nonumber\\
\varphi_{i}^{(2)}&=&\frac{3}{4}\varphi_{i}^{n}+\frac{1}{4}\varphi_{i}^{(1)}+\frac{1}{4}\Delta t L_{i}(\varphi_{i}^{(1)}) \nonumber\\
\varphi_{i}^{(n+1)}&=&\frac{1}{3}\varphi_{i}^{n}+\frac{2}{3}\varphi_{i}^{(2)}+\frac{2}{3}\Delta t L_{i}(\varphi_{i}^{(2)})
\end{eqnarray}
\\

\subsection{Flow Solver}
\begin{figure*}[ht!]
  \centering         
  \subfloat[Variables on a staggered mesh]{\label{fig:stag_mesh}\includegraphics[width=0.32\textwidth]{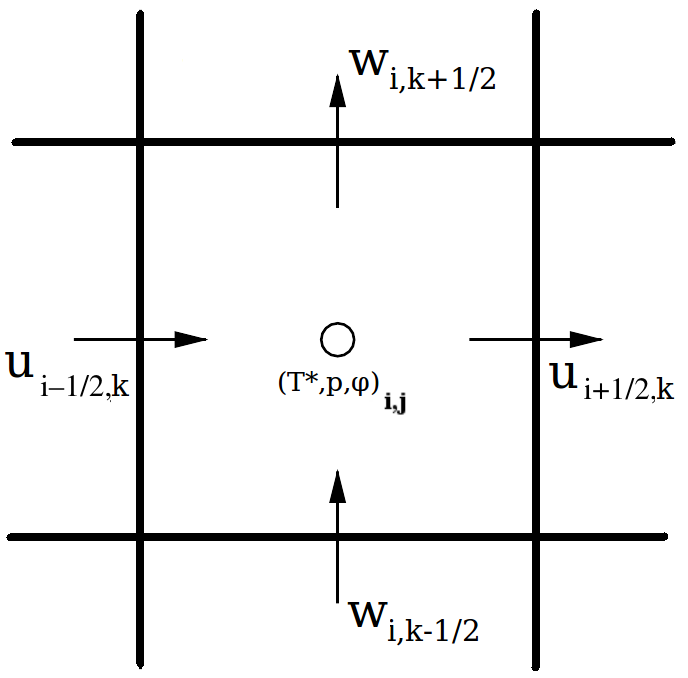}}
  \subfloat[Dual sub-mesh refined by factor R = 2]{\label{fig:stag_mesh2}\includegraphics[width=0.32\textwidth]{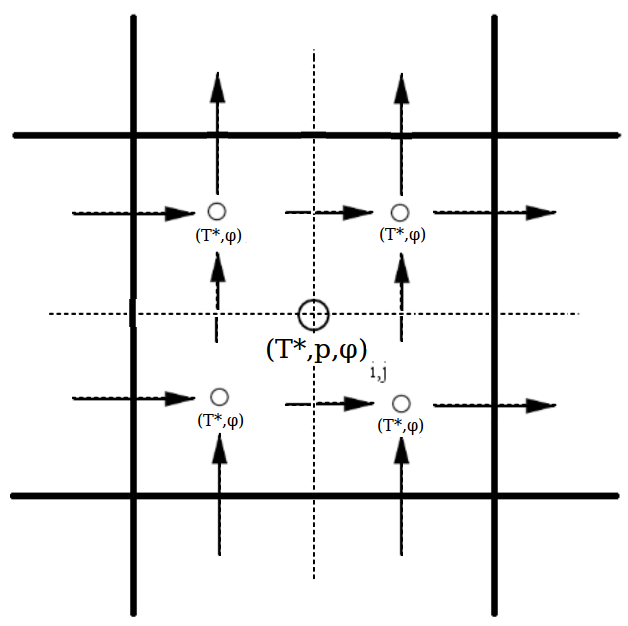}} 
  \subfloat[Dual sub-mesh refined by factor R = 3]{\label{fig:stag_mesh3}\includegraphics[width=0.32\textwidth]{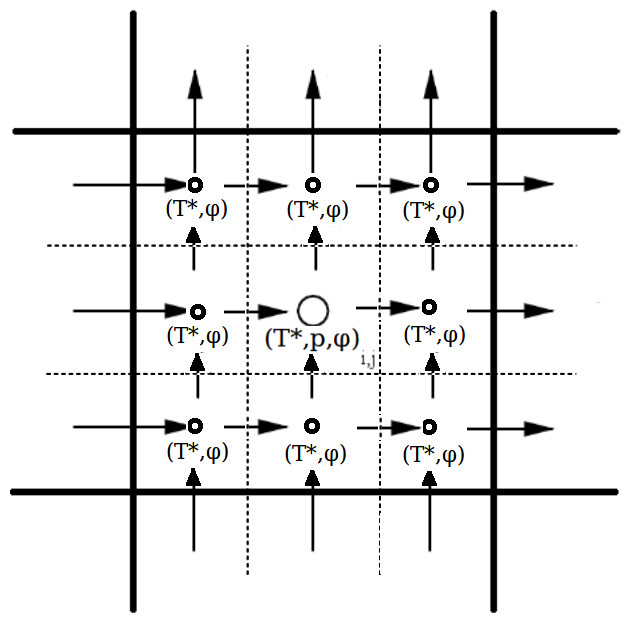}} 
  \caption{Variables on the new dual mesh. The flow field is solved on the outer coarse mesh, whilst the scalar is computed
on a refined subgrid. For the transport of the scalar the velocities are interpolated onto the midpoints of the subgrid}
  \label{fig:meshes}
\end{figure*}

This section outlines the numerical method of the flow solver used in the two-dimensional simulations presented in Sections 4 and 5. %buoyancy driven flow.
The velocity field is solved by a finite-difference discretization of the convective terms using a fourth-order
unconditionally kinetic energy conserving method combined with a fourth-order accurate central method for the diffusive terms \cite{Wissink_2004}. The 2D incompressible Navier-Stokes equation is
discretized on a non-uniform, staggered mesh in combination with a second-order accurate Adams-Bashforth time integration.
%The Boussinesq approximation is applied in order to account for the effects of buoyancy. 
The continuity equation (\ref{eqn:continuity}) for the two-dimensional incompressible flow 
%\begin{equation}
% \frac{\partial u}{\partial x}+\frac{\partial w}{\partial z}=0,
%\end{equation}
%which 
in discretized form on a mesh as shown in Fig.~\ref{fig:stag_mesh} 
reads
\begin{equation}
 \frac{u_{i+\frac{1}{2},k}^{n+1}-u_{i-\frac{1}{2},k}^{n+1}}{x_{i+\frac{1}{2}}-x_{i-\frac{1}{2}}}+
 \frac{w_{i,k+\frac{1}{2}}^{n+1}-w_{i,k-\frac{1}{2}}^{n+1}}{z_{k+\frac{1}{2}}-z_{k-\frac{1}{2}}}=0 
\end{equation}
%The momentum equations are given by
%\begin{eqnarray}
% \frac{\partial u}{\partial t}&=&-\frac{\partial p}{\partial x}
% + a \label{eqn:Naw_stoke_x} \\
% \frac{\partial w}{\partial t}&=&-\frac{\partial p}{\partial z}
% + c.  \label{eqn:Naw_stoke_y}
%\end{eqnarray}
%where $p$ is pressure and $a$ and $c$ represent the sum of the convective and diffusive terms
%\begin{eqnarray}
%a&=&-\frac{\partial u^{2}}{\partial x}-\frac{\partial uw}{\partial z}
% + \frac{1}{Re}\left\{\frac{\partial u^{2}}{\partial x^{2}}+\frac{\partial u^{2}}{\partial z^{2}}\right\} \label{eqn:a_momentum} \\
%c&=&-\frac{\partial w^{2}}{\partial z}-\frac{\partial uw}{\partial x}
% + \frac{1}{Re}\left\{\frac{\partial w^{2}}{\partial x^{2}}+\frac{\partial w^{2}}{\partial z^{2}}\right\}+\beta (T^*). 
% \label{eqn:c_momentum}
%\end{eqnarray} 
%where $Re=100$, based on a characteristic length scale of $L=1cm$ and a characteristic velocity of $u=w=1cm/s$
%The buoyancy term $\beta (T^*)$ in equation \eqref{eqn:c_momentum} is a function of the non-dimensional temperature $T^*$ 
%(see~\eqref{eqn:temp_nondimensionalise}) and is modelled using
%the Boussinesq approximation. In discretized form $\beta (T^*)$ reads,
%\begin{equation}
% \beta (T^*) = -0.5611516 \frac{T^*_k+T^*_{k+1}}{2}.
%\end{equation}

%The temperature difference between the top and
%the initial bulk temperature is only 3 \textcelsius. The temperature range in this particular case is between 20 and 23 \textcelsius \space where
%the relation between density and temperature can be assumed to be linear. 

When substituting the momentum equation into the continuity equation a Poisson equation for the pressure is obtained. 
The Poisson equation is iteratively solved using the conjugate gradient method with a diagonal preconditioning. From the
obtained pressure field the new velocity field can be calculated by rearranging the discretized equations of \eqref{eqn:Naw_stoke_x} and \eqref{eqn:Naw_stoke_y},
\begin{eqnarray}
 u^{n+1}_{i+\frac{1}{2},k}=u^{n}_{i+\frac{1}{2},k} + \varDelta t \left\{- \frac{p_{i+1,k}^{n+1}-p_{i,k}^{n+1}}{x_{i+1}-x_{i}}+a^{n}_{i+\frac{1}{2},k} \right\}  \\
 w^{n+1}_{i,k+\frac{1}{2}}=w^{n}_{i,k+\frac{1}{2}} + \varDelta t \left\{- \frac{p_{i,k+1}^{n}-p_{i,k}^{n+1}}{z_{k+1}-z_{k}}+c^{n}_{i,k+\frac{1}{2}} \right\}. 
\end{eqnarray}

Fig.~\ref{fig:stag_mesh} shows the location of variables on a staggered mesh. To achieve kinetic energy conservation
interpolations are required to evaluate the convective term. For instance at $(x_{i+\frac{1}{2}},z_{k})$ only the
$u$-velocity component is available at that location, while the $w$-velocity is only available at $(x_{i},z_{k+\frac{1}{2}})$.
Hence an interpolation of $w$ to the position where $u$ is defined gives $\overline{w}_{i+\frac{1}{2},k}$. An equivalent 
procedure for the $z$-momentum where $u$ needs to be interpolated where $w$ is defined gives $\overline{u}_{i,k+\frac{1}{2}}$. 
This yields to the discretization of the convective terms by a fourth-order central discretization,
\begin{align}
C_{x}(u,\overline{w})_{i+\frac{1}{2},k}=&
     - \frac{1}{2}\left[\frac{1}{-x_{i+\frac{5}{2}}+8x_{i+\frac{3}{2}}-8x_{i-\frac{1}{2}}+x_{i-\frac{3}{2}}} 
     \Big\{ -u_{i+\frac{5}{2},k}(u_{i+\frac{1}{2},k}+u_{i+\frac{5}{2},k})+8 u_{i+\frac{3}{2},k}(u_{i+\frac{1}{2},k}+u_{i+\frac{3}{2},k})  \right. \nonumber \\
     &\left. - 8 u_{i-\frac{1}{2},k}(u_{i+\frac{1}{2},k}+u_{i-\frac{1}{2},k})+u_{i-\frac{3}{2},k}(u_{i+\frac{1}{2},k}+u_{i-\frac{3}{2},k})\Big\} 
     + \frac{1}{-z_{k+2}+8z_{k+1}-8z_{k-1}+z_{k-2}}  \right.\nonumber  \\
    &\left.  \times \Big\{ -u_{i+\frac{1}{2},k+2}(\overline{w}_{i+\frac{1}{2},k}+\overline{w}_{i+\frac{1}{2},k+2})+8 u_{i+\frac{1}{2},k+1}(\overline{w}_{i+\frac{1}{2},k}+\overline{w}_{i+\frac{1}{2},k+1})   
     - 8 u_{i+\frac{1}{2},k-1}(\overline{w}_{i+\frac{1}{2},k}+\overline{w}_{i+\frac{1}{2},k-1}) \right.\nonumber  \\
    &\left. +u_{i+\frac{1}{2},k-2}(\overline{w}_{i+\frac{1}{2},k}+\overline{w}_{i+\frac{1}{2},k-2})\Big\}\right]
\end{align}
The convective terms in the $z$~-direction are discretized in a similar manner. The diffusive terms are discretized using the fourth-order accurate
central discretization scheme (\eqref{eqn:diffusion_discretize_x} and \eqref{eqn:diffusion_discretize_z}) in which the coefficients of the seven point stencil employed for the discretization on a non-uniform mesh are determined using Lagrange interpolations. 

\subsection{Dual Mesh Approach}
\label{sec:hybrid}
Because the diffusivity of the scalars of interest is up to three orders of magnitude smaller than that of the momentum, the resolution 
requirements for the flow field is less stringent as shown by the mesh refinement test in Section \ref{sec:Flow_field}. Thus, to save computing 
time a dual mesh approach is used as illustrated in Fig. \ref{fig:meshes}. The velocity is solved on a coarser base mesh (Fig.~\ref{fig:stag_mesh}), while 
the scalar is defined on the finer subgrid (Fig.~\ref{fig:stag_mesh2} and~\ref{fig:stag_mesh3}) so that the required computational resources are significantly reduced. 
  
To calculate the convective transport of the scalar, the velocities are 
interpolated onto a subgrid using a fourth-order Lagrange interpolation. When employing a subgrid refinement by a factor of $R = 2$ (Fig.~\ref{fig:stag_mesh2}) 
an interpolation is required for each subcell as the velocity location and the scalar locations do not coincide with their counter parts on the base mesh. 
In contrast, Fig.~\ref{fig:stag_mesh3} shows that in the case of a subgrid refinement by a factor $R=3$ some velocities and the
central subcells for a scalar are defined at the same locations.   
\subsection{Implementation of Boundary Conditions}
\label{sec:BoundaryConditions}
Dirichlet and Neumann boundary conditions are implemented by extrapolating 
the values obtained at the latest time step to ghost cells outside of the 
computational domain. This has the advantage that there is no need to change the
numerical stencils near boundaries. Suppose the quantity $q$ is defined on an
$N$-point mesh and we want to implement a Dirichlet (odd) boundary condition 
$q_N = Q$. By using the known values 
$q_{\frac{1}{2}},q_{\frac{3}{2}},\ldots,q_{N-\frac{1}{2}}$, the values 
$q_{N+i-\frac{1}{2}}$ are determined by using the formula
\begin{eqnarray}
q_{N+i-\frac{1}{2}}&=& 2\,Q -
q_{N-i+\frac{1}{2}}\;\;\mbox{for}\;\;i=1,\ldots,3.  \label{extraDirichlet}
\end{eqnarray}
To implement the Neumann (even) boundary condition at $i=0$, we use the formula
\begin{eqnarray}
q_{-i+\frac{1}{2}}&=&
q_{i-\frac{1}{2}}\;\;\mbox{for}\;\;i=1,\ldots,3.  \label{extraNeumann}
\end{eqnarray}
The free-slip condition for the velocity is implemented by using a Neumann
boundary condition~(\ref{extraNeumann}) of the velocity component that is parallel to the boundary
and a Dirichlet boundary condition~(\ref{extraDirichlet}) for the component that is orthogonal to 
the boundary (using the value zero at the boundary itself).
\section{1D Numerical Experiments}
\label{sec:1DNumericalExperiments}
In this section we apply the WENO-scheme for different test problems with the purpose of predicting the accuracy of the method 
on uniform and stretched meshes, respectively. As the problem is a convection-diffusion problem the numerical scheme was tested for both, scalar transport by convection only and by diffusion only.

%$\delta x_i = \frac{1}{2}(x_{i+\frac{1}{2}}-x_{i-\frac{1}{2}})$
\subsection{Scalar transport by convection on a uniform grid}
By using the modified quadratic Lagrange interpolations for 
reconstruction (\ref{eqn:P_lagrange}) we expect 
to achieve a fifth-order accuracy for the convective scalar transport on 
uniform meshes. 
In both of the following cases (uniform and non-uniform meshes), we use the previously described WENO schemes for spatial discretization and the $3^{rd}$-order 
Runge-Kutta-scheme for time integration of the one-dimensional convection equation. Because we want to predict the 
numerical error in the WENO scheme, physical diffusion will be neglected. If $\varphi(x_i,t)$ and $\varphi_{exact}$ are the 
numerical and the exact solutions, respectively at $(x_i,t)$, the $\L_1$ discretization 
error is given by
\begin{equation}
 \L_1= \frac{1}{N}\displaystyle\sum_{i=1}^{N}\mid\varphi(x_i,t)-\varphi_{exact}\mid, %\cdot \delta x_i  
\end{equation}
where $N$ describes the number of nodes in the domain, $t$ the time, $i$ the node number.  %and 

At first the WENO schemes are tested in a one-dimensional (1D) domain using uniform meshes. 
Therefore, the 1D scalar convection equation,
%\vspace{\baselineskip}
\begin{equation}
  \frac{\partial \varphi}{\partial t} + \frac{\partial \varphi}{\partial x} = 0, 
\end{equation}
was discretized on $0 \leq x \leq 2$ using periodic boundary conditions at $x=0$ and $x=2$. 
%\vspace{\baselineskip}
The scalar distribution was initialized by a sine wave function 
$\varphi(x,0) = \varphi_0(x)=\sin{(\pi x)}$. 

In the calculations an extremely small CFL-number was used so that the time-step would be 
small enough to ensure that the third-order temporal behaviour of the Runge-Kutta scheme would 
not affect the rate of convergence of the WENO schemes.

Table~\ref{tab:1d_uniform_test1} 
\begin{table}[ht]
\begin{center}
\begin{tabular}{|ccc|ccc|ccc|}
\hline
\multicolumn{3}{|l|}{WENO-\citet{Liu_osher_chan_94}} & \multicolumn{3}{|l|}{WENO-\citet{Jiang_Shu_96}} & \multicolumn{3}{|l|}{Upstream Central}\\
$N$ & $\L_1$-error & order  & $N$ & $\L_1$-error & order & $N$ & $\L_1$-error & order\\
\hline
%%%eps=10^-6
 10  & 1.17 E-02 & -    &  10  & 2.11 E-02 & -     &  10  & 3.11 E-03 & -    \\
 20  & 2.47 E-03 & 2.24 &  20  & 1.10 E-03 & 4.27  &  20  & 1.01 E-04 & 4.95 \\
 40  & 3.30 E-04 & 2.90 &  40  & 3.26 E-05 & 5.07  &  40  & 3.18 E-06 & 4.99 \\
 80  & 2.53 E-05 & 3.70 &  80  & 9.98 E-07 & 5.03  &  80  & 9.99 E-08 & 4.99 \\
 160 & 1.57 E-06 & 4.01 & 160  & 3.12 E-08 & 5.00  & 160  & 3.15 E-09 & 4.99 \\
 320 & 6.13 E-08 & 4.68 & 320  & 9.76 E-10 & 5.00  & 320  & 1.03 E-10 & 4.94 \\
 640 & 1.04 E-09 & 5.89 & 640  & 3.13 E-11 & 4.96  & 640  & 4.26 E-12 & 4.59 \\
%%%eps=10^-9
% 10  & 1.17 E-02 & -    &  10  & 2.11 E-02 & -     &  10  & 3.11 E-03 & -    \\
% 20  & 2.47 E-03 & 2.24 &  20  & 1.10 E-03 & 4.27  &  20  & 1.01 E-04 & 4.95 \\
% 40  & 3.30 E-04 & 2.29 &  40  & 3.26 E-05 & 5.07  &  40  & 3.18 E-06 & 4.99 \\
% 80  & 2.54 E-05 & 3.70 &  80  & 9.98 E-07 & 5.03  &  80  & 9.99 E-08 & 4.99 \\
% 160 & 1.64 E-06 & 3.95 & 160  & 3.12 E-08 & 5.00  & 160  & 3.15 E-09 & 4.99 \\
% 320 & 1.01 E-07 & 4.02 & 320  & 9.78 E-10 & 4.99  & 320  & 1.03 E-10 & 4.94 \\
% 640 & 5.97 E-09 & 4.08 & 640  & 3.16 E-11 & 4.95  & 640  & 4.26 E-12 & 4.59 \\
\hline
\end{tabular}
\end{center}
\caption{Absolute error and order of convergence on uniform meshes with $\varepsilon=10^{-6}$.}
\label{tab:1d_uniform_test1}
\end{table}
gives the $\L_1$ error after running the simulation during one time-unit as well as the resulting order of accuracy. The WENO5 implementation of \citet{Liu_osher_chan_94} was compared to the alternative WENO5 scheme developed by \citet{Jiang_Shu_96} and the upstream central method that is obtained by selecting the smoothness indicators $IS_i=0$ in either of the WENO schemes. Starting from $N=10$ nodes the $\L_1$ error is decreasing when increasing the number of nodes to 20, 40,..., 640. 
As previously found by \citet{Jiang_Shu_96}, the implementation of \citet{Liu_osher_chan_94} shows a smaller error than the scheme of \citet{Jiang_Shu_96} on the coarse 10-point mesh while on finer meshes the \citet{Jiang_Shu_96} implementation is superior. 

Furthermore, the scheme of \citet{Jiang_Shu_96} as well as the upstream central scheme show a fifth-order behaviour, while the original scheme of \citet{Liu_osher_chan_94} would need an even finer mesh to exhibit this behaviour. With the mesh sizes shown in the table, we would need to increase $\varepsilon$ significantly (even up to a value of $\varepsilon=1$) to achieve higher order. The choice of the small $\varepsilon=10^{-6}$ was necessary for the present application in order to resolve very steep gradients.
To test whether the \citet{Liu_osher_chan_94} scheme has the potential to exhibit a fifth-order behaviour, an additional test had been carried out in which $\varepsilon$  was increased to $\varepsilon=1$ (Table \ref{tab:1d_uniform_test1eps1}).
\begin{table}[ht]
\begin{center}
\begin{tabular}{|ccc|}
\hline
\multicolumn{3}{|l|}{WENO-\citet{Liu_osher_chan_94}} \\
$N$ & $\L_1$-error & order\\
\hline
10 & 3.46E-03 &  - \\
20 & 1.76E-04 & 4.30E+00\\
40 & 2.83E-06 & 5.96E+00\\
80 & 9.44E-08 & 4.91E+00\\
160& 3.11E-09 & 4.92E+00\\
320& 1.02E-10 & 4.93E+00\\
640& 4.26E-12 & 4.59E+00\\
\hline
\end{tabular}
\end{center}
\caption{Absolute error and order of convergence on uniform meshes with $\varepsilon=1$.}
\label{tab:1d_uniform_test1eps1}
\end{table}
As can be seen in Table \ref{tab:1d_uniform_test1eps1} for $\varepsilon=1$, indeed a fifth order behaviour for the original scheme was observed for $N \ge 80$ grid points. 
The slight decrease in order of convergence for $N=640$ points is possibly caused by machine-accuracy limitations affecting the calculations.

In practical calculations the mesh will be relatively coarse so that the original WENO implementation of \citet{Liu_osher_chan_94}, which has a good accuracy on coarse meshes, would be a good choice. 
Though the fifth-order upstream central method is shown to be even more accurate on coarse meshes, it is not the method of choice as the absence of a mechanism to deal with steep gradients could result in the appearance of wiggles as will be briefly discussed in Section \ref{sec:comp2D_WENO}.

%%%
\subsection{Scalar transport by convection on non-uniform meshes}
%%%

Using the modified Lagrange interpolations the WENO-scheme has been applied on non-uniform meshes on which the node distribution is given by:
% 
% \vspace{\baselineskip}
% 
% \begin{eqnarray}
%   x(i) = \tan{((i-0.5\cdot N) h)}. & 0 \leq i \leq N
% \end{eqnarray}
% where $h =\frac{2}{N}$. The mesh is then scaled so that $x(0)=0$ and $x(N)=2$. 
% 
% \vspace{\baselineskip}
% 
% The same conditions for calculating the errors and distribute the scalar in the domain have been used as for the uniform mesh. 
% Due to the variable node distances  $\Delta x_i$ caused by the stretching, the CFL-number changes within the domain. However 
% the maximal CFL-number has been set to $CFL_{max}=0.5$. 
% \begin{table}[ht]
% \begin{center}
% \begin{tabular}{|c|c|c|c|c|}
% \hline
% $N$ & $\L_1$-error & $\L_1$-order \\
% \hline
%  40  & 3.70 E-03 & -   \\
%  80  & 5.30 E-04 & 2.8 \\
%  160 & 7.60 E-05 & 2.8 \\
%  320 & 1.04 E-05 & 2.9 \\
% \hline
% \end{tabular}
% \end{center}
% \caption{Absolute error and order of convergence on non-uniform meshes.}
% \label{tab:1d_nonuniform_test1}
% \end{table}
%The node distribution in the z-direction from $z(0)=0$ to $z(n_{z})=5 L$ is then given by,
\begin{eqnarray}
x(i)&=&\left[ 1-\frac{\mathrm{tanh}(x_{\phi})}{\mathrm{tanh}(x_{1})}\right] x(0) + \left[ \frac{\mathrm{tanh}(x_{\phi})}{\mathrm{tanh}(x_{1})}\right] x(n_x)
\label{eqn:stretching1}
\end{eqnarray}
%with \begin{eqnarray}
%\omega&=&\frac{tanh(z_{\phi})}{tanh(z_{1})}
%\end{eqnarray}
for $i=1,...,n_x-1$, with 
\begin{eqnarray}
 x_{\phi}&=&\delta/2 \frac{i}{n_{x}} \nonumber
 \\
 x_1&=&\delta/2. \nonumber
\end{eqnarray}
The procedure for the stretching is controlled by the parameter $\delta$. 
The $N$-point mesh distribution is so that $x(0)=0$ and $x(n_{x})=1$ where 
$n_x=N/2$. The resulting mesh is subsequently mirrored about $x=1$ to obtain 
the grid points between $x(n_x)=1$ and $x(N)=2$.

The results of the tests using $\delta=1.0$ and $3.0$ is presented in  Table~\ref{tab:1d_nonuniform_test1}.
\begin{table}[h]
\begin{center}
\begin{tabular}{|ccc|ccc|}
\hline
\multicolumn{3}{|l|}{$\delta=1.0$} &\multicolumn{3}{|l|}{$\delta=3.0$} \\
$N$ & $\L_1$-error & $\L_1$-order & $N$ & $\L_1$-error & $\L_1$-order\\
\hline
10  & 1.20E-02 & -        &  10 & 3.58E-02 & -\\
20  & 2.41E-03 & 2.32E+00 &  20 & 6.28E-03 & 2.51E+00\\
40  & 3.91E-04 & 2.63E+00 &  40 & 1.60E-03 & 1.98E+00\\
80  & 6.27E-05 & 2.64E+00 &  80 & 3.77E-04 & 2.08E+00\\
160 & 1.41E-05 & 2.15E+00 & 160 & 9.26E-05 & 2.03E+00\\
320 & 3.45E-06 & 2.03E+00 & 320 & 2.31E-05 & 2.00E+00\\
640 & 8.62E-07 & 2.00E+00 & 640 & 5.77E-06 & 2.00E+00\\
\hline
\end{tabular}
\end{center}
\caption{Absolute error and order of convergence on non-uniform meshes with $\varepsilon=10^{-6}$.}
\label{tab:1d_nonuniform_test1}
\end{table}
 The absolute errors, as expected, are smaller for the mesh with reduced stretching. Compared to uniform meshes it can be seen that the order of accuracy decreases to approximately 2.                                                                    

\subsection{Scalar transport: pure diffusion}
In this section we apply the fourth-order accurate central discretization \eqref{eqn:diffusion_discretize_x} 
for the solution of scalar diffusion. 
In the one-dimensional case, concentration gradients in the $y$- and $z$~-directions are zero,
and we have the one-dimensional diffusion equation for a scalar $\varphi(x,t)$ 
\begin{eqnarray}
  \frac{\partial \varphi}{\partial t} = D\frac{\partial^{2} \varphi}{\partial x^{2}}. 
\label{eqn:diff_only}
\end{eqnarray}

For the test a one-dimensional domain was chosen with $0 \le x \le 5 L$. A mesh with $N$ grid points was defined with a refinement near the surface where the 
concentration boundary layer will form. At $x=0$ the boundary 
condition $\varphi(0,t) = 1$ was
imposed. The analytical solution for this boundary value problem
is given by
\begin{eqnarray}
  \varphi(x,t)=1-\mathrm{erf}\left(\frac{x}{\sqrt{4Dt}}\right).
\label{eqn:diff_analytical}
\end{eqnarray}
 The initial condition for the test was given by the analytical solution as defined in (\ref{eqn:diff_analytical}) at $t$=10 seconds. In the case of diffusive gas transfer into a liquid $D=\frac{1}{ReSc}$ and for the transfer of oxygen into water we have a Schmidt number of $Sc=500$ and a Reynolds number of 
$Re=100$, which is based on a characteristic length scale of $L=1$ cm and a characteristic velocity of $u=1$ cm/s. The latter gives us a characteristic time scale of $\theta=L/U=1$ second. 

The absolute errors and order of accuracy for the pure diffusion scalar transport on non-uniform meshes were tested for $N=10$ to $640$. The results after 1 time-unit are shown in Table \ref{tab:1d_nonuniform_diffusion}. 
\begin{table}[h]
\begin{center}
\begin{tabular}{|ccc|ccc|}
\hline
\multicolumn{3}{|l|}{$\delta=3.0$} &\multicolumn{3}{|l|}{$\delta=4.5$} \\
$N$ & $\L_1$-error & $\L_1$-order & $N$ & $\L_1$-error & $\L_1$-order\\
\hline
10  & 1.47E-04 & -         &  10 & 1.27E-03 & -\\
20  & 2.22E-04 & -5.95E-01 &  20 & 3.26E-04 & 1.96E+00\\
40  & 2.45E-04 & -1.41E-01 &  40 & 2.53E-05 & 3.69E+00\\
80  & 2.11E-05 &  3.54E+00 &  80 & 1.50E-06 & 4.07E+00\\
160 & 1.73E-06 &  3.61E+00 & 160 & 1.00E-07 & 3.91E+00\\
320 & 1.12E-07 & 3.95E+00  & 320 & 6.72E-09 & 3.90E+00\\
640 & 7.19E-09 & 3.96E+00  & 640 & 4.37E-10 & 3.94E+00\\
\hline
\end{tabular}
\end{center}
\caption{Absolute error and order of convergence on non-uniform meshes for pure diffusion case.}
\label{tab:1d_nonuniform_diffusion}
\end{table}
The absolute errors in the numerical results are very small, illustrating very good agreement with the analytical solution. 
A fourth order accuracy is achieved even with increased stretching.
% The test was carried out for various number of nodes $n_x=32, 64, 128$ and $ 288$. 
% Even with only 32 grid points the diffusion is resolved with an accuracy better than 1\%. 
% Figure~\ref{fig:difftest} shows a comparison of numerical and theoretical results of the concentration boundary
% layer after $t=80~L/U$. Note that this plot only shows every $4^{th}$ gridpoint in the region near to the boundary where the concentration gradient occurs.
% \begin{figure}[h!]
%  \centering
%  \includegraphics[width=0.55\textwidth, trim = 0mm 0mm 0mm 8mm, clip]{Diffusion_comparison_Sc500_80sec_skip4}
%  \caption{Comparison of numerical and analytical results of the diffusion concentration boundary layer}
%  \label{fig:difftest}
% \end{figure} 

All 1D numerical tests described above (Sections \ref{sec:1DNumericalExperiments}.1 to \ref{sec:1DNumericalExperiments}.3)  illustrate the advantageous behaviour of the chosen combination of a WENO-scheme 
with a fourth-order discretization of the diffusive terms which resulted in a low numerical diffusion and small 
absolute errors for both modes of transport, pure convection and pure diffusion, respectively.

\section{Two dimensional sheared scalar distribution}%Mesh sensitivity in 2D 
\label{sec:Mesh_sensitivity_2D_bolb}

\begin{figure}[h!] \centering
\subfloat[$t=$ 0 s]{\includegraphics[angle=-90,width=0.45\textwidth, trim = 5mm 5mm 5mm 5mm, clip]{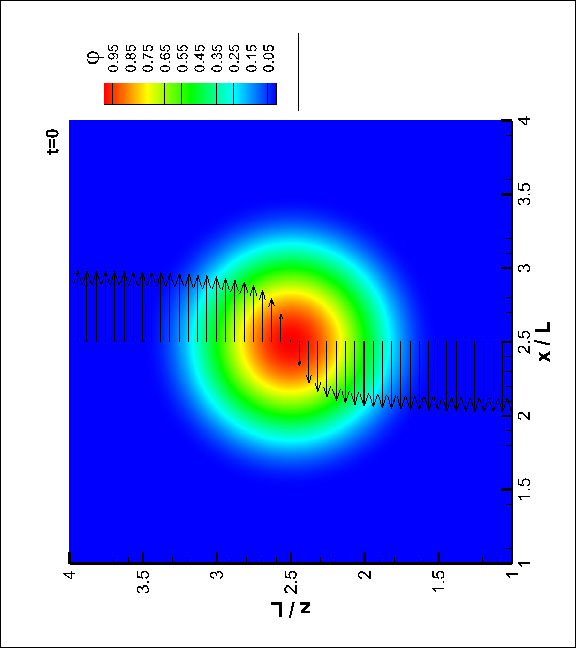}}
\subfloat[$t=$ 1.1 s]{\includegraphics[angle=-90,width=0.45\textwidth, trim = 5mm 5mm 5mm 5mm, clip]{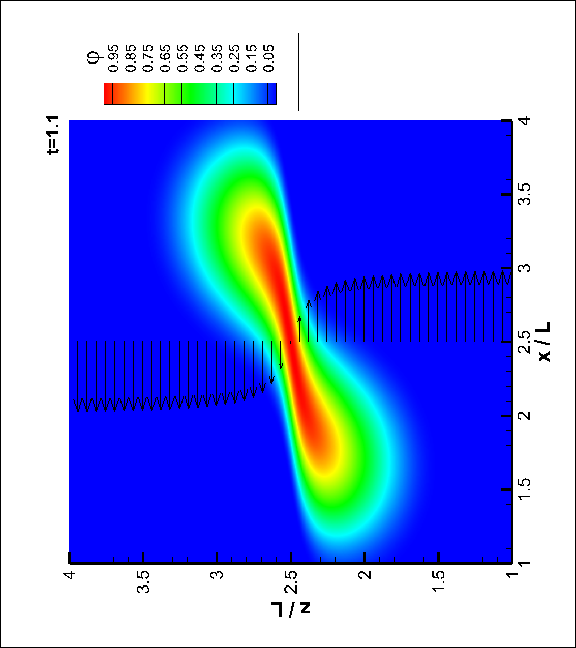}}
 \caption{A detail of the sheared scalar distribution. a) at $t$=0 second and b) at $t$=1.1 seconds.}
 \label{fig:sheared_scalar}
\end{figure}

To further test the robustness of the numerical scheme, we perform mesh sensitivity tests in 2D for two application cases, namely for sheared scalar distribution and low-diffusivity scalar transport in buoyancy driven flow. The first 
problem deals with a smooth scalar distribution without scalar diffusion being sheared by a zero viscosity flow as shown in Figure~\ref{fig:sheared_scalar}.  
After $1$ time-unit the flow is reversed with the aim to
obtain the initial distribution of the scalar back so that the distribution at $t=0$ should be the same as at $t=2$. 

The simulation was run on a $5L \times 5L$ domain using periodic boundary conditions in the 
horizonal direction and free-slip boundary conditions for the velocity combined with 
zero-flux boundary conditions~(\ref{extraNeumann}) for the scalar along the 
upper and lower boundaries. At $t=0$, the scalar field was initialised by 
\begin{equation}
\varphi_{i,k} = 0.5 \left (1 + \cos( \pi \sqrt{(x_i - 2.5)^2 + (z_k -  2.5)^2} \;) \right ),
\end{equation}
while the flow field was initialised using
\begin{equation}
 u_{i+\frac{1}{2},k} =  2\; \frac{\mathrm{atan}(10(z_k-2.5))}{\pi}.
\end{equation}
At $t = 1$ second the flow field was reversed, so that 
\begin{equation}
 u_{i+\frac{1}{2},k} = - 2\; \frac{\mathrm{atan}(10(z_k-2.5))}{\pi}.
\end{equation}
After $t = 2$ seconds of simulation the error is determined by comparing the initial to the calculated scalar distribution. 
As can be seen in Table \ref{tab:2d_shearedscalardistribution}, a grid refinement study was carried out by performing simulations on a sequence of uniform meshes with $80 \times 80$ up to $640 \times 640$ points. With increasing number of grid points the order of accuracy  was found to increase significantly from about 2 to 4.
\begin{table}[ht]\centering
\begin{tabular}{|ccc|}
\hline
$n_x \times n_z$   &  $\L_1$-error & $\L_1$-order \\
\hline
$40\times40$   &  1.34E-03  & - \\ 
$80\times80$   &  3.51E-04  & 1.93E+00 \\
$160\times160$ &  7.30E-05  & 2.26E+00 \\
$320\times320$ &  9.33E-06  & 2.97E+00 \\
$640\times640$ &  5.92E-07  & 3.98E+00 \\
\hline
\end{tabular}
\caption{Absolute error and order of convergence resulting from the 2D sheared scalar distribution test on uniform meshes using the WENO scheme of \citet{Liu_osher_chan_94} with $\varepsilon=10^{-6}$.}
\label{tab:2d_shearedscalardistribution}
\end{table}

\section{Two-dimensional low-diffusivity scalar transport in buoyancy driven flow} %Mesh sensitivity in 2D
\label{sec:Mesh_sensitivity_2D_buoyancy}
The second 2D application case considers the problem of low-diffusivity (high Schmidt number) mass transfer in buoyant-convectively driven flow. An example in nature is the oxygen absorption through the air-water interface in lakes at nighttime when the lakes' surface is cooled by the overlying cold air leading to unstable stratification which in turn causes mixing at the water side. 

\begin{figure}[h!]
 \centering
 \includegraphics[width=0.6\textwidth]{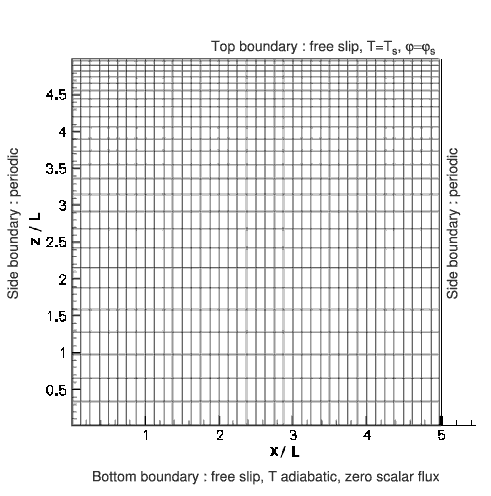}
 \caption{Schematic representation of the computational domain. The scatter shows every $10^{th}$ grid line of the major grid used for 
the velocity field. The mesh for the scalar was further refined by factors of $R=2, 3$ and $5$.}
 \label{fig:domain}
\end{figure}

The description of the 2D numerical setup for the problem is as follows. A quadratic domain was chosen with an edge length of $5 L$ as illustrated in Fig.~\ref{fig:domain} 
%illustrates a schematic representation of the computational domain. 
The base grid size was $n_{x}=400$ and $n_{z}=256$ in the $x$- and $z$-directions, 
respectively. The mesh was stretched in the $z$-direction with $\delta= 3$ to obtain a finer 
resolution near the top where a steep 
concentration gradient occurs. 
The general stretching procedure has been given in (\ref{eqn:stretching1}). 
%, the domain
% size is given by 
% \begin{eqnarray}
%  z(0)&=&0  \\
%  z(n_{z})&=&5 \nonumber 2
% \end{eqnarray}
% For $k=1,..., n_{z}-1$ 
% \begin{eqnarray}
%  z_{\phi}&=&(\delta/2)\times k/n_{z} \\
%  z_1&=&\delta/2  \nonumber
% \end{eqnarray}
% 
% Subsequently we have,
% \begin{eqnarray}
% \omega&=&\frac{tanh(z_{\phi})}{tanh(z_{1})}
% \end{eqnarray}
% The node distribution in the z-direction from $z(0)=0$ to $z(n_{z})=5 L$ is then given by,
% \begin{eqnarray}
% z(k)&=&(1-\omega)\times z(0) + \omega \times z(n_z)
% \end{eqnarray}
% for $k=1,...,n_z-1$.
For all variables, periodic boundary conditions were employed in the 
horizontal direction. For the velocity field free-slip 
boundary conditions were used at the top and bottom of the computational domain.
At the beginning of each simulation all velocity components were set to zero. 
The full set of 2D equations for the velocity given 
in~(\ref{eqn:continuity}) to (\ref{eqn:c_momentum}) is solved. It should be 
noted that to account for the effects of buoyancy in our application case 
the buoyancy term $\beta (T^*)$ is included into equation~\eqref{eqn:c_momentum} so that $c$ reads 
\begin{eqnarray}
c&=&-\frac{\partial w^{2}}{\partial z}-\frac{\partial uw}{\partial x}
 + \frac{1}{Re}\left\{\frac{\partial^2 w}{\partial x^{2}}+\frac{\partial^2 w}{\partial z^{2}}\right\} +\beta (T^*). 
 \label{eqn:c_momentumbuoy}
\end{eqnarray}
The term $\beta (T^*)$ is modelled using the Boussinesq approximation and is a 
function of the non-dimensional temperature $T^*$ defined as 
\begin{eqnarray}
  %T^{*}=\frac{T-T_0}{T_B-T_0}.
  T^{*}=\frac{T-T_s}{T_{B,0}-T_s}
\label{eqn:temp_nondimensionalise}
\end{eqnarray}
where the temperature at the top of the domain was set to a fixed value of  
$T=T_s=20$\textcelsius \space, see~(\ref{extraDirichlet}), while in the rest of the computational domain 
the initial bulk temperature $T_{B,0}$ 
was 23\textcelsius. \space The relation between density and temperature within 
this range can be assumed to be linear. To avoid heat losses, at the bottom of
the computational domain an adiabatic boundary condition (\ref{extraNeumann}) was employed for $T$. The temperature $T$ is a scalar and hence treated the same as $\varphi$, see (\ref{eq:conv_diff}). 

At the top of the computational domain the scalar $\varphi$ was kept at a 
value of $\varphi = \varphi_s$ (\ref{extraDirichlet}) while at the bottom a zero-flux boundary condition (\ref{extraNeumann}) was
employed. The scalar was non-dimensionalized using the scalar magnitude at 
the top boundary $\varphi_{s}$ and the initial magnitude in the bulk 
$\varphi_{B,0}=0$ so that
\begin{eqnarray}
  %\varphi^{*}=\frac{\varphi-\varphi_B}{\varphi_0-\varphi_B}
  \varphi^{*}=\frac{\varphi-\varphi_{B,0}}{\varphi_s-\varphi_{B,0}}.
\label{eqn:phi_nondimensionalise}
\end{eqnarray}

The convective instability was triggered by adding random disturbances to the temperature 
field after letting it evolve for $t=11$ seconds in order to avoid the triggering of the instability to depend on the mesh size or numerical round off error. 
The same disturbance field was used in all simulations. 
The random numbers that were added to $T^*$ were uniformly distributed between 
$0$ and $T_{ran}$. To test the influence of the level of the random disturbances on the development 
of the instability, a test was performed in which a random disturbance field was rescaled so that 
$T_{ran}=0.010$, $T_{ran}=0.020$ and $T_{ran}=0.040$ before it was added 
to the non-dimensional temperature. 
In all three simulations exactly the same buoyant convective disturbance field was found to develop. 
As can be seen in Table~\ref{tab:randomdisturbance}, 
\begin{table}[h!]\centering
\begin{tabular}{|cc|}
\hline
$T_{ran}$ &   time at which the falling plume reaches z=4.0 cm \\
\hline
0.010  &                                             23.75 s \\
0.020  &                                             22.30 s \\
0.040  &                                             20.85 s  \\
\hline
\end{tabular}
\caption{The time difference found between the development of disturbances.}
\label{tab:randomdisturbance}
\end{table}
the different levels of disturbances were found to affect the time 
it takes for the plumes to develop.
Based on the time difference of 1.45 seconds between subsequent simulations (in which the level of random disturbances is doubled) the exponential growth factor $\lambda$ for the buoyant-convective instability was estimated to be $\lambda=0.478$.
\par
To facilitate the comparison between various simulations involving buoyant convection, 
in the simulations discussed below (with the exception of Section~\ref{sec:experiments}) 
the same random temperature field consisting of uniformly distributed random numbers between $T=0$ and $T_{ran}=0.020$ 
was added to the non-dimensional temperature field. 
\subsection{Comparison of scalar convection methods in 2D} \label{sec:comp2D_WENO}
As mentioned briefly in Section 3.1 although the fifth-order upstream central method (C5) shows better accuracy on coarse meshes, 
it is not the method of choice for the current application due to the absence of a mechanism to deal with steep gradients which could result in the appearance 
of wiggles. To demonstrate this we performed a number of initial 2D simulations on the $400 \times 256$ base mesh using the C5 and the WENO5 schemes discussed in Section 3. 
\begin{figure}[h]
 \centering
 \includegraphics[width=0.6\textwidth, trim = 0mm 0mm 0mm 0mm, clip]{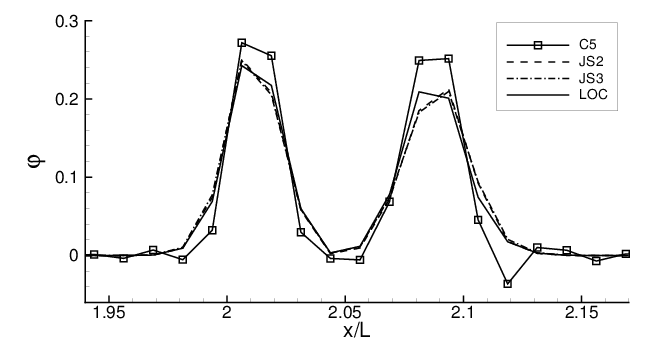}
  \caption{Comparison of WENO5 schemes (JS2, JS3, LOC) and the fifth-order 
central scheme (C5), showing profiles of the scalar distribution $\varphi$ at $z=4.5$cm and $t=45$ seconds using a Schmidt number of $Sc=500$.}
 \label{fig:wiggles}
\end{figure} 
Figure \ref{fig:wiggles} shows the profiles extracted at a cross section at $z=4.5$ cm and $t=45$ seconds 
obtained for the C5 scheme and different variants of the WENO5 scheme. The cross-section intersects with the falling plumes that develop due to the convective instability which induces sharp gradients in the scalar distribution (see also Figure~\ref{fig:45s_standard}).
The plot reveals that wiggles appear close to steep gradients when using the C5 method.
It was found that the wiggles completely disappear when we use the WENO5 schemes JS2 and JS3 of \citet{Jiang_Shu_96} with powers $r=2$ and $3$, respectively - see~(\ref{eq:aplus_JiangShu},\ref{eq:aminus_JiangShu}) - 
as well as the original implementation of \citet{Liu_osher_chan_94} (LOC). It can be seen that the results obtained using the WENO5 schemes are very similar. 
\par
In the following sections, the mesh sensitivity for resolving the 2D flow and concentration fields were tested in several subgrid mesh refinement studies. 

\subsection{Mesh sensitivity test : Flow-field} \label{sec:Flow_field}
\begin{figure*}[ht!]
  \centering         
  \subfloat[vertical velocity $w$ field on a grid $400 \times 256$ and $800 \times 512$ after $t=45~$seconds]{\label{fig:w_vel_vertical}\includegraphics[width=0.36\textwidth]{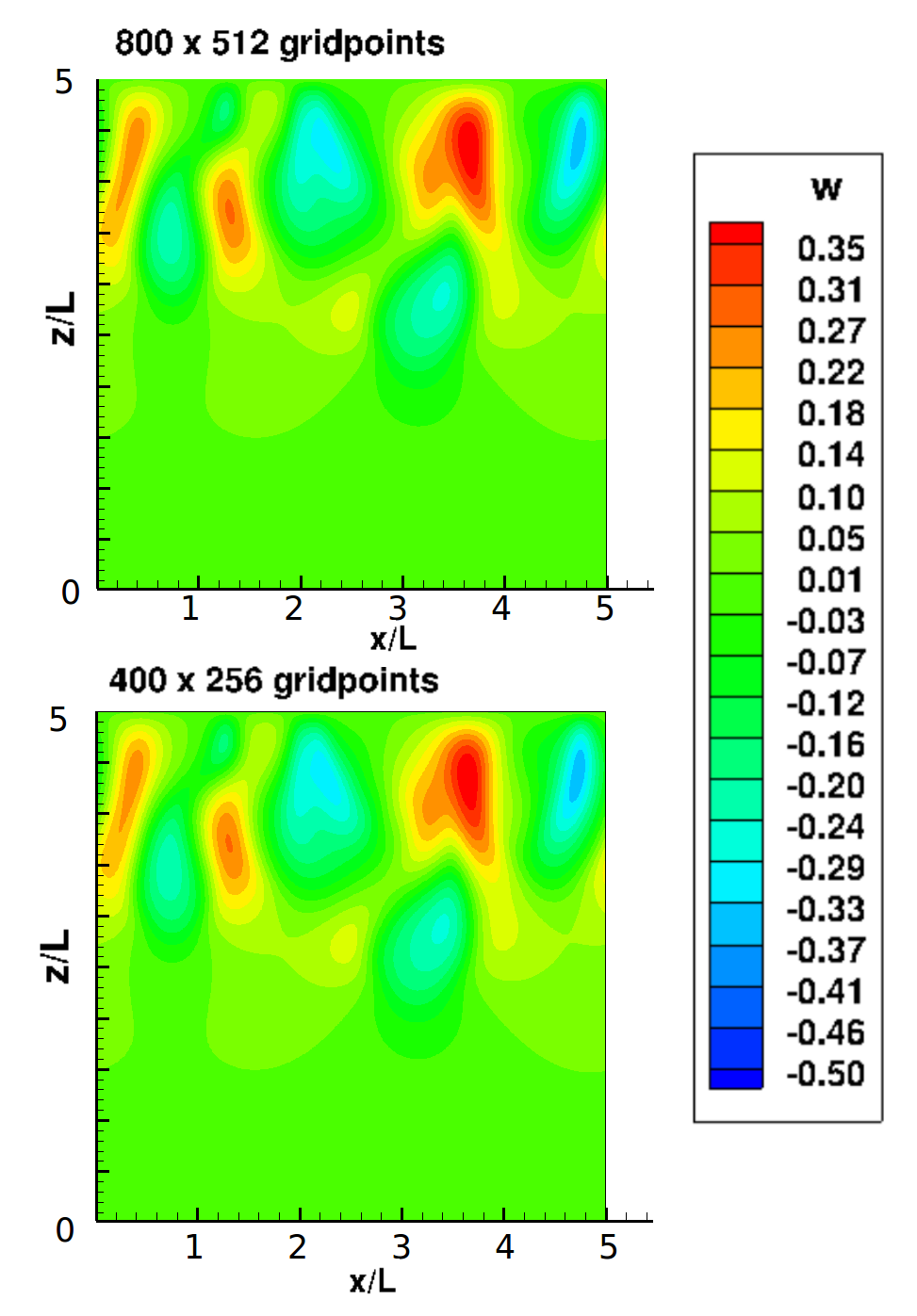}}
  \subfloat[vertical velocity $w$ along a line at $z=4L$ after $t=45~$seconds]{\label{fig:w_vel_plots}\includegraphics[width=0.55\textwidth, trim = 0mm 0mm 0mm 15mm, clip]{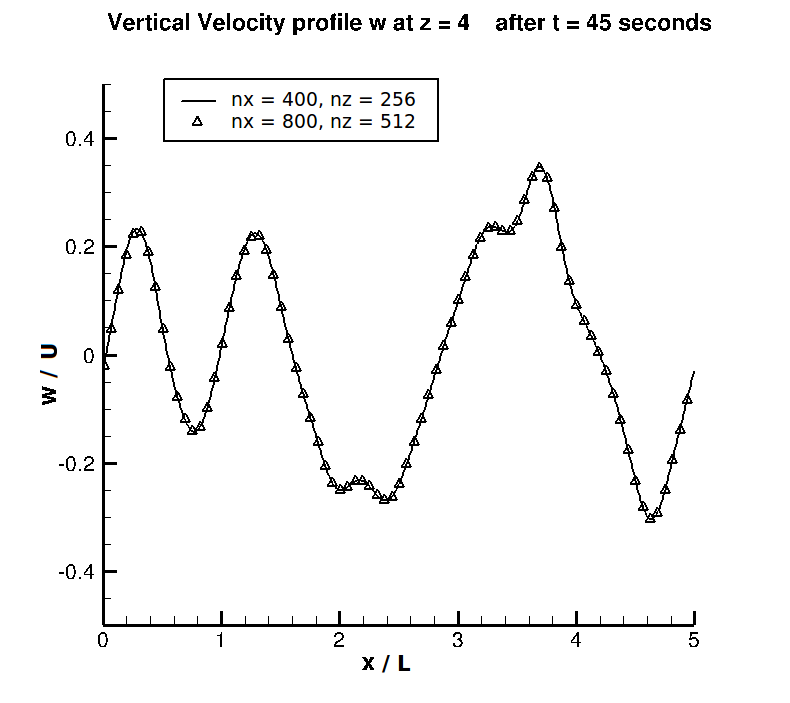}}  \\
%  \subfloat[Scalar $\varphi$ for $Sc =500$ at $z = 4.7L$]{\label{fig:45s_z_4p7}\includegraphics[width=0.2\textwidth]{Sc500_55sec}} 
%  \subfloat[Scalar $\varphi$ for $Sc =500$ at $z = 4.6L$]{\label{fig:45s_z_4p6}\includegraphics[width=0.5\textwidth]{lineplots_4p6_45s_2}}\\
  \caption{A grid refinement showed that the velocity field is fully resolved on a $400 \times 256$ grid.}
  \label{fig:w_vel_comparsion_400_800}
\end{figure*}

To verify that the flow-field was fully resolved on the chosen $400 \times 256$ base mesh, 
the grid was refined in all directions by factors of 1.5 and 2, respectively. Fig.~\ref{fig:w_vel_comparsion_400_800} 
shows the contour plots and the velocity profile obtained 
from the simulations with the base grid and the mesh refined by a factor 2 ($R=2$) 
(with $800 \times 512$ points) after $t=45~$seconds. The contour plots of the flow-field using the refined mesh did not show any
visible changes in the flow structures (Fig.~\ref{fig:w_vel_plots}). 
This is further confirmed by the vertical velocity profiles along a horizontal line at $z=4 L$. The profiles show a nice convergence verifying that the velocity field 
is fully resolved on the $400 \times 256$ grid which was subsequently used in all further cases. 

\subsection{Mesh sensitivity test : gas concentration field}
\begin{figure*}[ht!]
  \centering         
  \subfloat[Domain after t =  45 seconds on standard mesh]{\label{fig:45s_standard}\includegraphics[width=0.5\textwidth, trim = 0mm 0mm 0mm 19mm, clip]{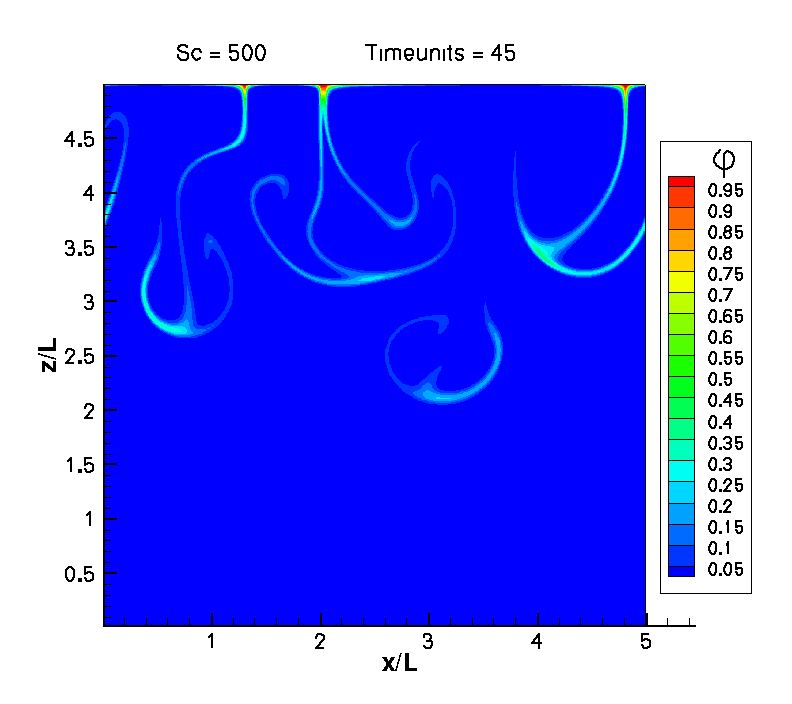}}
  \subfloat[Domain after t =  45 seconds with refined submesh $R=3$]{\label{fig:45sec_hybrid}\includegraphics[width=0.5\textwidth, trim = 0mm 0mm 0mm 19mm, clip]{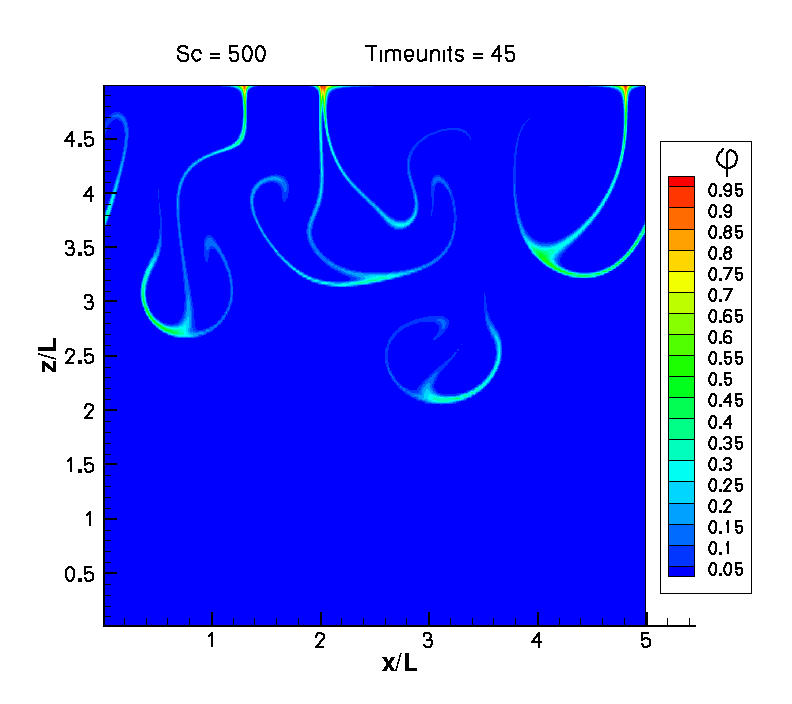}} \\
  \subfloat[Zoomed view after t = 45 seconds on standard mesh]{\label{fig:45s_standard_zoom}\includegraphics[width=0.5\textwidth, trim = 0mm 0mm 0mm 19mm, clip]{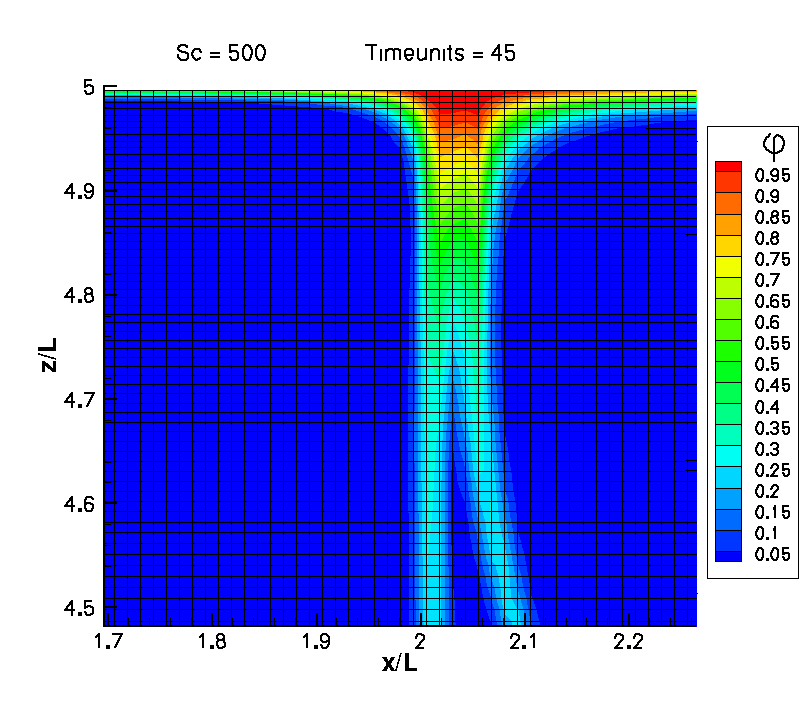}}
  \subfloat[Zoomed view after t = 45 seconds with refined submesh $R=3$]{\label{fig:45s_hybrid_zoom}\includegraphics[width=0.5\textwidth, trim = 0mm 0mm 0mm 19mm, clip]{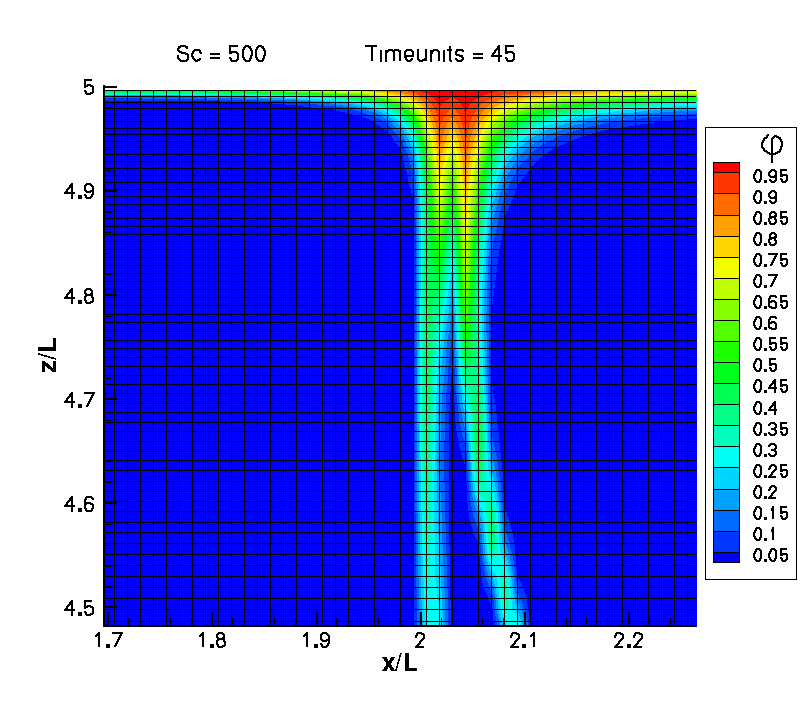}}\\
  \caption{Comparison of the gas concentration field after t =  45 seconds on standard mesh $n_{x}=400$ and $n_{z}=256$ and with a dual
 subgrid in place three times as fine (see Fig.~\ref{fig:stag_mesh3}). The gas concentration field is resolved in sharper
detail with less smearing.}
  \label{fig:flow_plots}
\end{figure*}

As described above, we used a dual-mesh approach in which the scalar was resolved on a finer mesh than the one used for the velocity. 
Various levels of refinement were employed as illustrated in Fig~\ref{fig:meshes}.            
In this section, the mesh sensitivity for the scalar transport  using this dual mesh approach is evaluated.  
Fig.~\ref{fig:flow_plots}
shows a comparison of the non-dimensional gas concentration contour plots that visualise the development
of the scalar transport at $t=45~$seconds using the base mesh ($400 \times 256$) for both velocity and scalar and the dual mesh approach with refinement factor 3
applied to the scalar. The Schmidt number is $Sc=500$ which is equivalent to
the diffusion of oxygen in water. 

In general, both concentration 
fields in Figs.~\ref{fig:45s_standard} and \ref{fig:45sec_hybrid} reveal the same structures of downwards plumes. However, a zoomed 
view of the top region near the water surface reveals a more detailed representation of the gas concentration 
field when the dual mesh approach is used (Fig.~\ref{fig:45s_standard_zoom} and~\ref{fig:45s_hybrid_zoom}). Please note that the gas concentration in all figures is interpolated to the base grid and not shown on the refined mesh used 
for the scalar transport.

\begin{figure*}[ht!]
  \centering         
  \subfloat[Scalar $\varphi$ for $Sc =500$ at $x = 2.0L$]{\label{fig:45s_x_2p0}\includegraphics[width=0.5\textwidth, trim = 0mm 0mm 0mm 28mm, clip]{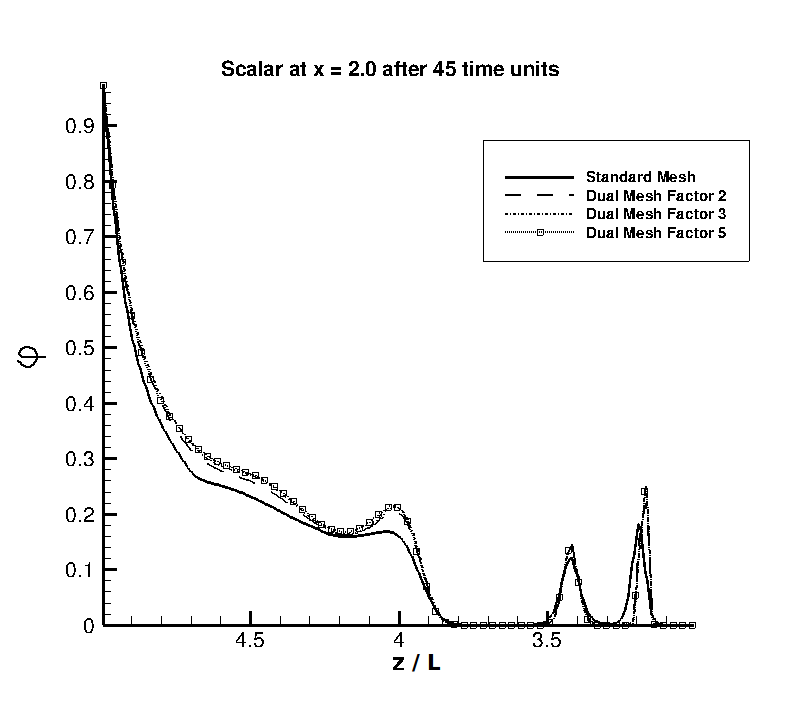}}
  \subfloat[Scalar $\varphi$ for $Sc =500$ at $x = 1.3L$]{\label{fig:45s_x_1p3}\includegraphics[width=0.5\textwidth, trim = 0mm 0mm 0mm 28mm, clip]{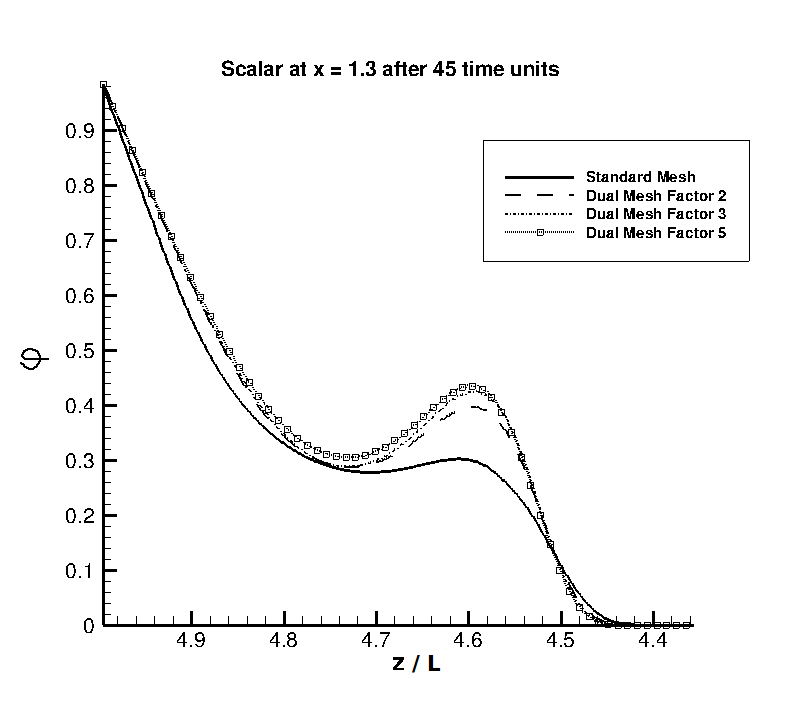}} \\
  \caption{Comparison of scalar field after t =  45 seconds on different levels of subgrid mesh refinement. The two locations are vertical lines  
at $x = 2.0L$ and $x=1.3L$ along the downwards plumes as seen on Fig.~\ref{fig:45s_standard}.}
  \label{fig:x_line_plots}
\end{figure*}

\begin{figure*}[ht!]
  \centering         
  \subfloat[Scalar $\varphi$ for $Sc =500$ at $z = 4.9L$]{\label{fig:45s_z_4p9}\includegraphics[width=0.5\textwidth, trim = 0mm 0mm 0mm 28mm, clip]{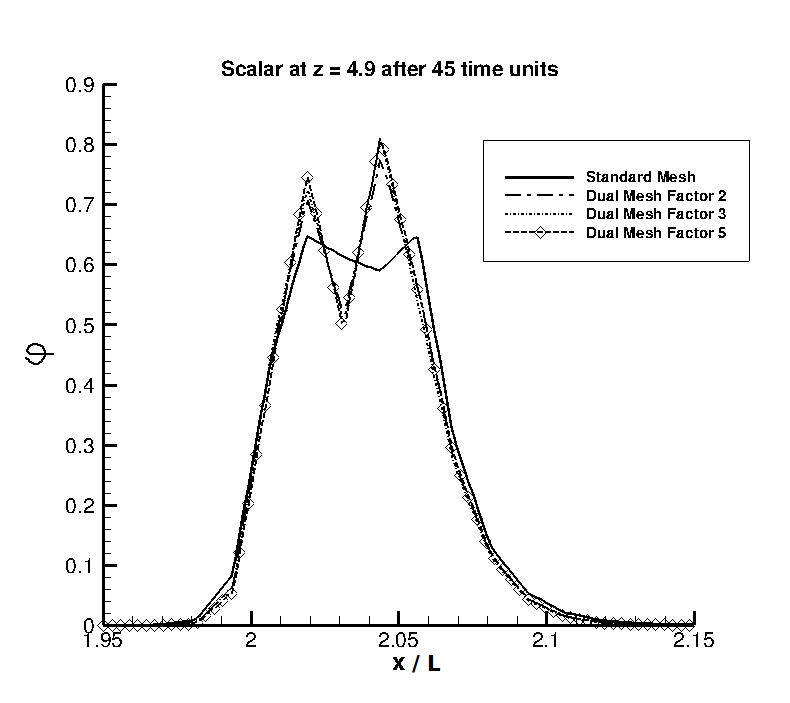}}
%  \subfloat[Scalar $\varphi$ for $Sc =500$ at $z = 4.8$]{\label{fig:45s_z_4p8}\includegraphics[width=0.5\textwidth, trim = 0mm 0mm 0mm 28mm, clip]{lineplots_4p8_45s_2}} \\
  \subfloat[Scalar $\varphi$ for $Sc =500$ at $z = 4.7L$]{\label{fig:45s_z_4p7}\includegraphics[width=0.5\textwidth, trim = 0mm 0mm 0mm 27.5mm, clip]{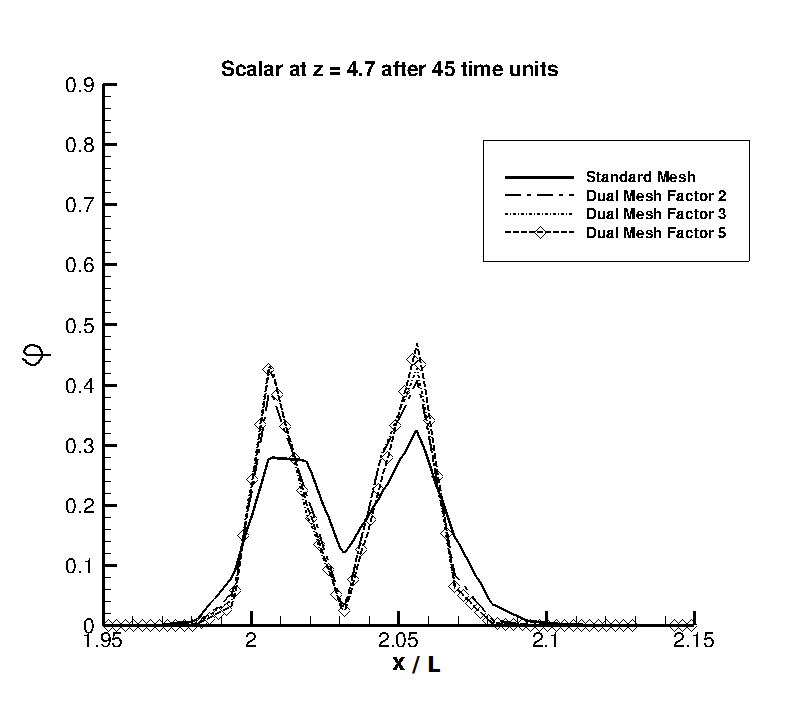}} \\
%  \subfloat[Scalar $\varphi$ for $Sc =500$ at $z = 4.6$]{\label{fig:45s_z_4p6}\includegraphics[width=0.5\textwidth, trim = 0mm 0mm 0mm 27.5mm, clip]{lineplots_4p6_45s_2}}
  \caption{Comparison of scalar field after t =  45 seconds on different levels of subgrid mesh refinement. The location is a horizontal line at various 
depths $z$ across the downwards plumes as seen on Fig.~\ref{fig:45s_hybrid_zoom} and~\ref{fig:45s_standard_zoom}.}
  \label{fig:z_line_plots}
\end{figure*}

Figs.~\ref{fig:x_line_plots} and ~\ref{fig:z_line_plots}
 show line plots  of the scalar field at various locations within the domain. The locations are across or along the typical mushroom pattern that develops as a result of the convective instability where 
sharp gradients in the scalar field are present. Solving the scalar on the finer subgrid shows a significant improvement in resolution.
The $R = 2$ refinement shows a big improvement in deeper regions where the base mesh is relatively coarse and the scalar distribution is maintained better. 
In the far field ($z < 4$) the scalar concentration profiles for the refined cases $R = 2$, $R = 3$ and $R=5$ converge to nearly
identical values (Fig.~\ref{fig:45s_x_2p0}).

\begin{figure*}[ht!]
  \centering         
  \subfloat[Scalar $\varphi$ for $Sc =500$ over
  time]{\label{fig:time_int_Sc_500}
  \includegraphics[width=0.5\textwidth, trim = 0mm 0mm 0mm 7mm, clip]{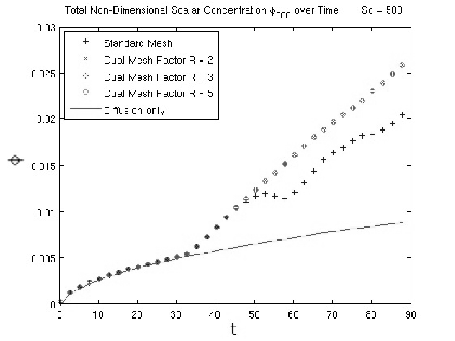}}  
  \subfloat[Non-dimensional total temperature $T^* $ over time]{\label{fig:time_int_temp}
  \includegraphics[width=0.5\textwidth, trim = 0mm 0mm 0mm 7mm, clip]{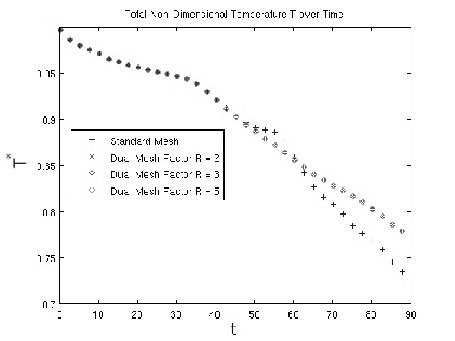}} 
%  \subfloat[Concentration field at $t=40$ and $t=55$ seconds]{\label{fig:45_55_Sc500_vertical}\includegraphics[width=0.26\textwidth]{Sc500_40sec_55sec_vertical}} \\
%  \subfloat[Scalar $\varphi$ for $Sc =500$ at $z = 4.7L$]{\label{fig:45s_z_4p7}\includegraphics[width=0.2\textwidth]{Sc500_55sec}} 
%  \subfloat[Scalar $\varphi$ for $Sc =500$ at $z = 4.6L$]{\label{fig:45s_z_4p6}\includegraphics[width=0.5\textwidth]{lineplots_4p6_45s_2}}\\
  \caption{Comparison of the total non-dimensional scalar concentration and temperature $T^* $over time for different levels of subgrid refinement}
  \label{fig:time_integration_Sc500_and_temp}
\end{figure*}

The improved resolution becomes even more relevant when the spatially integrated total scalar
concentration in the domain over time is considered. Fig.~\ref{fig:time_int_Sc_500}
shows the total concentration over time 
for $Sc =500$. Up to a time of $t=30 ~$seconds the gas transfer is dominated by diffusion. Subsequently, the instability 
induces a convective flow that enhances the mass transfer. The typical mushroom patterns start penetrating the deeper regions of the domain.
It is here where the refined submesh shows a much improved resolution with a continuous increase in the concentration levels whilst the 
standard mesh shows a drop in concentration levels. 
The drop occurs when the scalar reaches the region for $z<2.5 L$ (where the mesh
becomes significantly coarser) after around $t=55~$seconds (Fig.~\ref{fig:time_int_Sc_500}). This points out an insufficient resolution of the scalar transport in this region. This effect was not present 
in the refined cases (Fig.~\ref{fig:time_int_Sc_500}).
The same is found for the transport of the non-dimensionalized temperature $T^* $ (Fig. \ref{fig:time_int_temp}).
The grid refinement
study for the temperature transport shows a similar trend as seen for the concentration field in Fig.~\ref{fig:time_int_temp}. On the coarse mesh
fluctuations become evident after $t=50 ~$seconds whereas the refined cases do not exhibit such temperature fluctuations.
Again the results are identical for all refined cases ($R = 2$, $R = 3$ and $R=5$). 

\subsection{Comparison to Experiments} \label{sec:experiments}
In this section we compare the obtained scalar field with the laboratory 
measurements conducted by \citet{Jirka_Herlina_2010} at KIT. In the 
experiments instantaneous 2D  oxygen concentration fields were visualized 
using a Laser Induced Fluorescence (LIF) technique. %method based
%on the oxygen quenching phenomenon developed by Vaughan \& Weber %\cite{Vaughan_Weber_1970}. 
%The technique enabled an
%instantaneous 2D visualisation with good resolution of the dissolved oxygen concentration. 
The experiments were performed
in a $50\times50\times65$~cm$^3$ tank and the water depth was about $42$ cm. 
The surface temperature was 3 \textcelsius \hspace{1mm}  lower than the bulk 
temperature of the water. The equivalent temperature boundary condition was 
applied in the numerical simulations.

\begin{figure*}[ht!]
  \centering         
  \subfloat[Experimental
  Results]{\label{fig:experiment}\includegraphics[width=\textwidth, trim =
  0mm 0mm 11mm 0mm, clip]{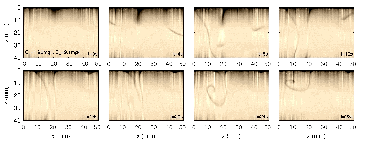}} \\
  \subfloat[Numerical Results]
  {\label{fig:numerical}\includegraphics[width=1.0\textwidth, trim =
  0mm 0mm 10mm 0mm, clip]{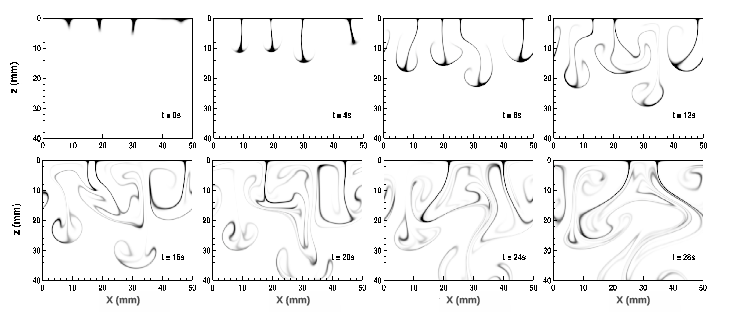}} 
  \caption{Comparison of flow structures. High oxygen concentration plumes of LIF measurements conducted by \citet{Jirka_Herlina_2010} (Fig.~\ref{fig:experiment}) and DNS results (Fig.~\ref{fig:numerical}). The dark and light colour scaling indicate regions with high and low scalar concentration, respectively. 
In both cases the surface temperature was 3\textcelsius \hspace{1mm} colder than the bulk temperature.}
  \label{fig:comparison}
\end{figure*}

Fig.~\ref{fig:comparison} 
shows a comparison of 2D-LIF images to 2D DNS results where a refinement factor of $R = 2$ has been used. Identical boundary conditions were employed as in the simulations described in the beginning of Section 5.
% The DNS calculations were performed using a $10L \times 10L$ computational domain and a $800 \times 299$ mesh that was stretched in the z-direction. 
% Identical boundary conditions were employed as in the simulations described in Section~\ref{sec:Mesh_sensitivity_2D_buoyancy}. The buoyant instability, however, was triggered using a different random field.
\par
Note that the DNS results show the 
top section of the domain that has the same dimension as the LIF-maps. The actual experimental domain was much larger so the sides
and bottom in these plots can be considered as open boundaries. For reasons of better comparison
the timescale was set to $t=0 ~$seconds from the moment
when the flow field started moving which was after a simulation time of 33 seconds. Though the numerical results are only 2D, whereas the real problem is of course 
3D, a very good qualitative agreement with the experiment is observed. 
%Especially between $t = 16 ~$seconds and
%$t = 28 ~$seconds 
Both the spatial distance between high concentration plumes and the size of the eddies were found to be similar in the
experiment and the simulation. Because of the low diffusivity of oxygen in water and the rather low turbulent flow the plumes of 
high oxygen concentration retain their fine structures. This means that the steep concentration gradients 
do not smear out because of excessive numerical diffusion. As a result good qualitative agreement between 
the numerical simulations and the experimental data is obtained. 

\section{Conclusion}

To accurately resolve the mass transport for a scalar with low diffusivity on a stretched and 
staggered mesh, the fifth-order accurate WENO5 schemes 
of \citet{Liu_osher_chan_94} and \citet{Jiang_Shu_96} was implemented to discretize 
the convective terms, while a fourth order accurate central 
discretization was used for the diffusion. The flow field was approximated by 
fourth-order accurate central discretizations for 
both convection and diffusion. Because the diffusivity of the scalars of interest is up to 
almost three orders of magnitude smaller 
than the molecular diffusivity of the ambient fluid, the resolution requirements for the 
transported scalar are much higher. 
Hence, to save computing time, a dual meshing approach was employed in which the scalar 
transport equations were discretized on a finer mesh than the flow field. 
The discretization of scalar convection and diffusion were tested in both 1D and 2D cases. 
The 1D tests showed that the spatial 
discretization of the scalar convection achieved a second-order accuracy on non-uniform meshes 
and a fifth-order accuracy on uniform meshes, while the discretization of the diffusive term 
was shown to achieve a fourth-order accuracy on stretched meshes. 
Though the WENO5 implementation of Jiang \& Shu \cite{Jiang_Shu_96} shows 
superior results in the grid refinement tests, the original scheme of Liu at al. \cite{Liu_osher_chan_94} was found to be more accurate on coarse meshes. 
Hence, to obtain a satisfactory resolution of the steep concentration gradients - that will occur 
in scalar transport problems with high Schmidt numbers - using as few grid points as possible, the Liu et al. implementation was found to be a good choice. 
\par
For the 2D 
case a combined active and passive scalar transport problem was simulated. It was shown that the fifth-order central upwind method generated wiggles near steep gradients which completely disappeared when using the WENO5 schemes. The dual meshing approach showed a significant improvement in
accuracy of the scalar field resolution even for a moderate refinement by a factor of two. Subsequent refinements 
that were carried out using factors of up to five times only showed marginal further improvements in accuracy. 
Additionally, a qualitative comparison of the numerical results to experimentally visualized oxygen concentration fields 
in water showed similar structures below the air-water interface even though the numerical simulations were 2-dimensional.

\bibliographystyle{plainnat}%{elsarticle-num}
\bibliography{references}

\end{document}